\documentclass[aip,amsmath,amssymb,reprint]{revtex4-1}
\usepackage{graphicx}
\usepackage[utf8]{inputenc}
\usepackage[T1]{fontenc}
\usepackage{mathptmx}
\usepackage[english]{babel}
\newcommand{\Rho}{\mathrm{P}}
\begin{document}
\preprint{AIP/123-QED}
\title{How antagonistic salts cause nematic ordering and behave like diblock copolymers}
\author{David Jung}
\affiliation{Forschungszentrum J{\"u}lich, Helmholtz Institute Erlangen-N{\"u}rnberg for Renewable Energy (IEK-11), F{\"u}rther Stra{\ss}e 248, 90429 N{\"u}rnberg, Germany}
\affiliation{Department of Theoretical Physics I, Friedrich-Alexander-Universit{\"a}t Erlangen-N{\"u}rnberg, N{\"a}gelsbachstra{\ss}e 49b, 91052 Erlangen, Germany}
\author{Nicolas Rivas}
\affiliation{Forschungszentrum J{\"u}lich, Helmholtz Institute Erlangen-N{\"u}rnberg for Renewable Energy (IEK-11), F{\"u}rther Stra{\ss}e 248, 90429 N{\"u}rnberg, Germany}
\affiliation{Millenium Nucleus Physics of Active Matter, Universidad de Chile, Blanco Encalada 2008, Santiago, Chile}
\author{Jens Harting}
\affiliation{Forschungszentrum J{\"u}lich, Helmholtz Institute Erlangen-N{\"u}rnberg for Renewable Energy (IEK-11), F{\"u}rther Stra{\ss}e 248, 90429 N{\"u}rnberg, Germany}
\affiliation{Department of Applied Physics, Eindhoven University of Technology, P.O. Box 513, 5600MB Eindhoven, The Netherlands}
\begin{abstract}
We present simulation results and an explanatory theory on how antagonistic salts affect the spinodal decomposition of binary fluid mixtures. We find that spinodal decomposition is arrested and complex structures form only when electrostatic ion-ion interactions are small. In this case fluid and ion concentrations couple and the charge field can be approximated as a polynomial function of the relative fluid concentrations alone. When the solvation energy associated with transfering an ion from one fluid phase to the other is of the order of a few $k_BT$, the coupled fluid and charge fields evolve according to the Ohta-Kawasaki free energy functional. This allows us to accurately predict structure sizes and reduce the parameter space to two dimensionless numbers. The lamellar structures induced by the presence of antagonistic salt in our simulations exhibit a high degree of nematic ordering and the growth of ordered domains over time follows a power law. This power law carries a time exponent proportional to the salt concentration. We qualitatively reproduce and interpret neutron scattering data from previous experiments of similar systems. The dissolution of structures at high salt concentrations observed in these experiments agrees with our simulations and we explain it as the result of a vanishing surface tension due to electrostatic contributions. We conclude by presenting 3D results showing the same morphologies as predicted by the Ohta-Kawasaki model as a function of volume fraction and suggesting that our findings from 2D systems remain valid in 3D.
\end{abstract}
\maketitle
\section{Introduction}
Antagonistic salts are defined by the special property that, in a mixture of high and low permittivity fluids, one constituent ion kind migrates towards regions of higher permittivity while the other ion type does the opposite. Recent experimental studies of the antagonistic salt NaBPh$_4$, solvated in a mixture of heavy water and 3-methylpyridine oil (3MP), suggest that adding antagonistic salt can cause structures to form on the nanometer scale~\cite{sadakane_membrane_2013}. These structures seem to be periodically repeating regions of high water/Na$^+$ concentration and high oil/BPh$_4^-$ concentration. Their length scale depends on many factors, such as the salt concentration~\cite{sadakane_membrane_2013}, volume fraction, and temperature~\cite{sadakane_multilamellar_2009}. Such mixtures show visible coloration with strong temperature sensitivity when the structure periodicity is of the order of the wavelength of visible light~\cite{sadakane_periodic_2007}. In other words, any fluid mixture of unequal permittivities may in principle be turned into a liquid crystal by simply adding an antagonistic salt. Being able to control mesoscopic structure formation in liquid crystals has potential for optical and nanoscale manufacturing technologies, such as directed self-assembly~\cite{muller_continuum_2018}.
\par
The theory of antagonistic salts has been studied to a notable extent by the group of Onuki~\cite{onuki_structure_2016}. Their model describes the onset of structure formation in a system near the critical temperature of fluid demixing and the characteristic length of these structures~\cite{onuki_solvation_2004}. Antagonistic salt mixtures have also been studied numerically by Araki, Okamoto and Onuki~\cite{araki_dynamics_2009,onuki_solvation_2011}, showing lamellar structure formation and various degrees of nematic ordering. Recently, Tasios et al.~\cite{tasios_microphase_2017} employed a simple lattice-based Monte Carlo algorithm to obtain a rich phase diagram of 3D structures from antagonistic salt mixtures. In the present work we add to this body of research an explicit analytical description of the coupling between fluid and ion concentrations and use it to show that antagonistic salt mixtures are very well approximated by the same free energy model commonly used to treat diblock copolymers~--~the Ohta-Kawasaki model~\cite{ohta_equilibrium_1986}. The Ohta-Kawasaki model is known to lead to ordered liquid crystal structures~\cite{zhang_periodic_2006,weith_stability_2013,cheng_efficient_2017} seemingly identical to the structures we obtain numerically from antagonistic salt systems. By approximating our system with the Ohta-Kawasaki model we can predict for which parameters there is strong coupling of fluid and ion concentrations, which leads to structure formation, and when these structures become unstable due to zero or negative surface tension resulting from electrostatic contributions. We also add a more quantitative treatment of the dynamics of nematic ordering, showing that nematically ordered domains grow continuously at a rate proportional to the salt concentration.\par
This article is structured as follows. Section~\ref{sec:method} explains our numerical method and justifies the simulation parameters used. We introduce a free energy model to derive the forces of fluid-ion coupling and demonstrate the relationship of our system to the Ohta-Kawasaki model. Our main results are split into five subsections. In Subsection~\ref{sec:coupling} we derive a polynomial function of the order parameter to approximate the charge distribution. Using this result we demonstrate in Subsection~\ref{sec:ohta} how our system is related to the Ohta-Kawasaki model. In Subsection~\ref{sec:nematic} we present our simulation results on nematic ordering. In Subsection~\ref{sec:lengths} we discuss and interpret experimental data from small-angle neutron scattering in light of our results and validate our model by comparing theoretical predictions of the characteristic length with the simulation results. Subsection~\ref{sec:3D} concludes with a few examples of the different morphologies possible in 3D depending on volume fractions. Finally, in Sec.~\ref{sec:outlook} we summarize our results and discuss possibilities for future research.
\section{Method}
\label{sec:method}
We use the lattice-Boltzmann method (LBM) in an implementation previously presented by Rivas et al.~\cite{rivas_mesoscopic_2018,rivas_solvation_2018}. The LBM is an efficiently parallelizable numerical method to model fluid dynamics which is equivalent to the Navier-Stokes equation whenever flow velocities are much smaller than the speed of sound. In the LBM, space is discretized in a grid of lattice sites $x_0$ apart, each containing a set of populations $f_i.$ The origins of the LBM in statistical physics lead to the interpretation of each population as proportional to the probability of finding a fluid particle travelling in the direction given by the index $i$, but the fluid is modeled purely as a continuum nonetheless~\cite{benzi_lattice_1992}. The number of populations per lattice site is determined by the coarseness of the velocity discretization. We use 19 populations per lattice site (commonly called the D3Q19 scheme), corresponding to a zero vector representing resting particles plus 6 velocity vectors connecting the lattice site to its closest neighbours plus 12 to the next further neighbours in the grid~\cite{qian_lattice_1992}. The populations are evolved in each time step by an advection step, in which populations are simply moved to the lattice site they are connected with by their corresponding velocity vector, and by a collision step, in which viscous energy dissipation and fluid forcing are accounted for by relaxing the populations to an equilibrium distribution. In the single-relaxation-time scheme by Bhatnagar, Gross and Krook~\cite{bhatnagar_model_1954} (BGK) that we use, the collision step takes the following shape:
\begin{equation}
  \label{eq:collision}
  f_i\longrightarrow f_i-\frac{1}{\tau}(f_i-f_i^\mathrm{eq})+S_i.
\end{equation}
$f_i^\mathrm{eq}$ is a discretized Maxwell-Boltzmann distribution expanded in terms of Hermite polynomials up to second order in velocity~\cite{kruger_lattice_2017,benzi_lattice_1992}.~$\tau$ gives the fraction of a time step, over which full relaxation to equilibrium takes place and determines the viscosity of the fluid.~$S_i$, called the source term, contains the modification of the local equilibrium state due to any fluid forces added to the system. Our chosen method to calculate the source term goes back to Guo et al.~\cite{guo_discrete_2002}. A mixture of two fluids is simulated by simply representing each fluid component $\sigma$ by a separate set of populations $f^\sigma_i.$ The local mass concentrations of our fluid components are then $n_\sigma=\sum_i f^\sigma_i.$ We treat both fluids equally, i.e.~they have equal density and viscosity.\par
The order parameter $\psi=(n_1-n_2)/(n_1+n_2)\in [-1,1]$ encodes the local fluid composition. We assume fluid incompressibility, i.e. $n_1+n_2=\mathrm{const.}=n_\Sigma.$ Fluid demixing is modeled by the pseudopotential method introduced by Shan and Chen~\cite{shan_lattice_1993}. In this method, an interaction parameter $G$ encodes the strength of repulsive forces between unlike fluid components. These repulsive forces are included in the respective fluid's source term $S^\sigma_i$ like any other force. For sufficiently high $G$, spinodal decomposition into two bulk phases occurs, whereas for sub-critical $G$ the phases mix homogenously.~$G$ encodes both material properties and the effect of temperature on intermixability.\par
Experimental evidence of D$_2$O/3MP mixtures shows structure formation induced by antagonistic salts at temperatures below the critical point, where macroscopic demixing would not occur without salt~\cite{sadakane_periodic_2007}. According to Onuki and Kitamura, this may be the effect of large thermal concentration fluctuations near the critical point of demixing stabilizing under the influence of the antagonistic salt~\cite{onuki_solvation_2004}. As our model does not include thermal fluctuations, we only study the effect of antagonistic salt at values of $G$ slightly above the critical point of demixing.\par
The electrokinetic forces acting on fluids and ions are derived from a free energy functional $\mathcal{F}_T$ including fluid demixing via a phenomenological Ginzburg-Landau term $\mathcal{F}_F$ and ionic contributions $\mathcal{F}_I$ modelling the salt as an ideal gas coupled to the fluid by a solvation potential with an electrostatic contribution from the ionic charges.
\begin{equation}
\begin{split}
  \label{eq:iongibbs}
    \mathcal{F}_T=\mathcal{F}_F+\mathcal{F}_I=\overbrace{\int g(\psi,\nabla\psi)}^{\mathcal{F}_F}+\\ \underbrace{\sum_\pm k_BT c_\pm \large[ \mathrm{ln}(v_0 c_\pm) -1 \large] + \sum_\pm c_\pm \mu_\pm^\mathrm{sol}(\psi)+\frac{1}{2}\rho\phi \hspace{1ex} \mathrm{d}V}_{\mathcal{F}_I}.
\end{split}
\end{equation}
Here, $c_\pm$ is the concentration of positive/negative ions, $v_0$ is the ionic volume (assumed equal for both ion types for simplicity), $\rho=e(z_+c_++z_-c_-)$ is the charge concentration with the elementary charge $e$ and valence $z_\pm$, and $\phi$ is the electrostatic potential satisfying $\nabla^2 \phi=-\rho/\epsilon.$ For simplicity, we assume the permittivity $\epsilon$ to be the same for both fluids and hence homogenous in space. We use a linearized solvation potential $\mu_\pm^{\mathrm{sol}}=\Delta \mu_\pm \psi /\Delta \psi$ with an ion-specific constant antagonicity $\Delta\mu_\pm$. $\Delta\mu_\pm$ is the solvation energy associated with transfering an ion from the bulk of one fluid phase to the other in a macroscopically demixed state where in the $n_1$-dominant phase $\psi=\psi_1$ and in the $n_2$-dominant phase $\psi=\psi_2$ fulfilling $\psi_1-\psi_2=\Delta\psi$. Note that throughout this work we write the Laplace operator as $\nabla^2$, while $\Delta$ signifies a variable related to a finite difference.\par
Similar models have been used in the past to theoretically describe binary fluid mixtures with antagonistic salts~\cite{onuki_ginzburg-landau_2006,rotenberg_coarse-grained_2009,rivas_mesoscopic_2018}. The fluid demixing term $\mathcal{F}_F=\int g(\psi,\nabla\psi)\hspace{0.5ex} \mathrm{d}V$ incorporates the surface tension $\gamma_{sc}$ caused by the repulsive pseudopotential forces in the LBM implementation. In order to model bulk demixing it should have the shape of a polynomial with two minima at the bulk values $\pm\Delta\psi/2$ of $\psi$ plus a gradient term. We choose $g(\psi,\nabla\psi)=g_0(\psi^2-\Delta\psi^2/4)^2+0.5\kappa_{sc}|\nabla \psi|^2.$~$g_0$ determines the energy penalty for mixed states. Its magnitude is irrelevant in our considerations. Minimization of $\mathcal{F}_F$ alone results in two bulk phases with an interface in the shape of a hyperbolic tangent $\psi(x)=0.5\Delta\psi \,\mathrm{tanh}\,(x/\lambda_I).$ We derive the interfacial width $\lambda_I=\Delta \psi^{-1} \sqrt{2 \kappa_{sc}/g_0}$ by demanding that the chemical potential $\delta \mathcal{F}_F/\delta \psi=0$ at equilibrium. The surface tension is estimated by integrating the interfacial energy density from bulk to bulk, i.e.~$\gamma_{sc}=0.5 \int_{-\infty}^{+\infty}\kappa_{sc}|\nabla \psi|^2\, \mathrm{d}x=\Delta\psi^3 \sqrt{\kappa_{sc} g_0/72}$, for a hyperbolic tangent shaped interface.\par
As long as spurious currents and inertial hydrodynamics are negligible, the pseudopotential model of demixing has been shown by Sbragaglia, Scarbolo et al.~to correspond to a similar monotonically decreasing free energy functional~\cite{sbragaglia_continuum_2009,scarbolo_unified_2013}. While full equivalence of the pseudopotential model to a free energy functional requires an additional gradient-shaped forcing term as a function of the order parameter, this term has been shown by numerical investigation to be generally negligible~\cite{sbragaglia_continuum_2009}.\par
The ion concentrations $c_\pm$ live on the same discrete lattice as the fluid and are evolved via a finite-difference scheme~\cite{rivas_mesoscopic_2018}. By using a mean-field approach to model ion dynamics we neglect ion pair interactions, which may become important at salt concentrations high enough to cause significant steric interactions. The simplest possible time evolution of $c_\pm$ to conserve ion numbers and minimize the free energy in Eq.~(\ref{eq:iongibbs}) is given by the Cahn-Hilliard equation~\cite{penrose_thermodynamically_1990}. In addition to the fluxes given by the Cahn-Hilliard equation we evolve the ions by an advection term in order to couple them to the velocity field $\vec{u}$ of the fluid mixture.
\begin{equation}
  \label{eq:cahnhilliard}
  \begin{aligned}
  \frac{\partial c_\pm}{\partial t}&=-\nabla \cdot \vec{J}_\pm -\nabla \cdot (\vec{u} c_\pm)\\
  \vec{J}_\pm&=c_\pm \mathcal{M} \nabla \mu_{c_\pm}.
  \end{aligned}
\end{equation}
$\mathcal{M}$ is the ion mobility, which can in general depend on the ion concentration and ion species. In order to recover Fick's laws of diffusion we set $\mathcal{M}=D/k_BT$ according to the Einstein-Smoluchowski relation, with an ion diffusivity $D$ assumed to be constant. By performing a functional derivative of the ionic contributions $\mathcal{F}_I$ to the free energy, we obtain an expression for the ionic chemical potential:
\begin{equation}
  \label{eq:chempotion}
  \mu_{c_\pm}=\frac{\delta \mathcal{F}_I}{\delta c_\pm}= k_BT \mathrm{ln}(v_0c_\pm)+ \frac{\Delta \mu_\pm}{\Delta \psi} \psi + ez_\pm \phi,
\end{equation}
where we use $\phi=e/4\pi\epsilon\int (z_+c_+-z_-c_-)/(|\vec{r}-\vec{r}_0|)\hspace{0.5ex} \mathrm{d}V$ to calculate the functional derivative of the electrostatic term. In solving Poisson's equation, as in general, we treat our system as 3D, but with a thickness of one lattice site and periodic boundary conditions in the third dimension when emulating a 2D system. Using Eq.~(\ref{eq:chempotion}) we can write out the total ion flux $\vec{J}_\pm$ from Eq.~(\ref{eq:cahnhilliard}). The resulting time evolution is identical to the well-known Nernst-Planck equation supplemented by a solvation term. The fluxes can be subdivided into diffusive, solvation, and electrostatic contributions $\vec{J}_\pm=\vec{j}^d_\pm+\vec{j}^s_\pm+\vec{j}^e_\pm$:
\begin{equation}
\begin{aligned}
\vec{j}^d_\pm&=-D\nabla c_\pm\\
\vec{j}^s_\pm&=-\frac{D}{k_BT}c_\pm \frac{\Delta\mu_\pm }{\Delta\psi}\nabla \psi\\
\vec{j}^e_\pm&=-\frac{D}{k_BT}z_\pm e c_\pm \nabla \phi.
\end{aligned}
\end{equation}
Because the fluid is friction-coupled to the ions, it experiences the same force density as the ions. Assuming the inertial time scale of the ions to be much smaller than that of the fluid we expect the fluxes to correspond to the instantaneously reached drift velocity of the ions times the ion concentration and derive the force density experienced by the ions from Stokes' law as
\begin{equation}
  \label{eq:ionforce}
  \vec{F}_{I\pm}=\frac{k_BT}{Dc_\pm}\vec{J}_\pm=- k_BT \nabla c_\pm -\frac{\Delta\mu_\pm}{\Delta \psi} c_\pm\nabla \psi -ez_\pm c_\pm \nabla \phi.
\end{equation}
In local equilibrium thermodynamics, spatial variations of intensive properties such as the chemical potential drive thermodynamic forces. The change in Gibbs free energy associated with a spatial variation of the chemical potential of fluid composition $\mu_\psi$ is $\mathrm{d}G=\psi \nabla \mu_\psi \mathrm{d}\vec{r}$, so that we can identify the corresponding force performing work on the system over an infinitesimal displacement $\mathrm{d}\vec{r}$ as
\begin{equation}
\begin{gathered}
  \label{eq:gibbs-duhem}
  \vec{F}_{fs}=-\psi \nabla \mu_\psi = - \sum_\pm \psi \frac{\Delta \mu_\pm}{\Delta \psi} \nabla c_\pm,\\
  \mu_\psi=\frac{\delta \mathcal{F}_I}{\delta \psi}= \sum_\pm c_\pm \frac{\Delta \mu_\pm}{\Delta \psi}.
\end{gathered}
\end{equation}
The second term from the right in Eq.~(\ref{eq:ionforce}) proportional to $\nabla \psi$ represents the migration of ions towards regions of high concentration of the fluid species they are preferably solvated by. The fluid experiences this ion solvation force due to the aforementioned friction-coupling to the ions. Additionally the fluid experiences an analogous fluid solvation force representing migration of the fluid towards regions of high concentration of the ion kind preferably solvated by the local fluid composition. This force is not friction-coupled back to the ions because of the unequal inertial time scales of fluid and ions. The ions indirectly experience this force via the advection term in Eq.~(\ref{eq:cahnhilliard}). In summary, the fluid forcing takes the form
\begin{equation}
\begin{split}
  \label{eq:fluidforce}
  \vec{F}_f=\vec{F}_{I+}+\vec{F}_{I-}+\vec{F}_{fs}=\\-\sum_\pm \big( \underbrace{k_BT \nabla c_\pm}_{\vec{F}_d} +\underbrace{\frac{\Delta\mu_\pm}{\Delta \psi} \nabla (c_\pm \psi)}_{\vec{F}_s} +\underbrace{ez_\pm c_\pm \nabla \phi}_{\vec{F}_e} \big).
\end{split}
\end{equation}
The equation of state, taking into account the gradient-shaped ideal ion pressure and total solvation forcing terms $\vec{F}_d$ and $\vec{F}_s$, but not the nonconservative electrostatic force $\vec{F}_e$, is then
\begin{equation}
  \label{eq:eqstate}
  p=\frac{x_0^2}{3t_0^2} (n_2+n_1)+\frac{x_0^2}{3} G \Psi_1\Psi_2+\sum_\pm \bigg( k_BT c_\pm+\frac{\Delta \mu_\pm}{\Delta\psi}c_\pm\psi \bigg).
\end{equation}
We write lattice units with subscript $0$, i.e.~the lattice length as $x_0$, one time step as $t_0$ and the units of mass as $m_0.$ In Eq.~(\ref{eq:eqstate}) we introduced the pseudopotential $\Psi_\sigma=m_0 x_0^{-3}(1-\mathrm{exp}(-x_0^3 n_\sigma/m_0)).$ It approximates the fluid concentration of component $\sigma$, but is limited to values $\leq m_0x_0^{-3}$ (1 in simulation units). It is used in place of the fluid concentrations $n_\sigma$ only in order to avoid numerical instabilities, where fluid compression might increase the pseudopotential forces, in turn causing further fluid compression. The factors of 3 in Eq.~(\ref{eq:eqstate}) stem from the speed of sound of $1/\sqrt{3}x_0t_0^{-1}$ dictated by the chosen D3Q19 scheme of spatial discretization. \par
We restrict ourselves to monovalent ions $z_+=-z_-=1$ and symmetric antagonism $\Delta \mu_+=-\Delta \mu_-=\Delta \mu$ from here on. Simulation parameters roughly corresponding to experimental values can be chosen in the following way. First, we choose a length scale suitable to resolve the expected structure lengths, e.g. $x_0\approx 1$nm. The unit of mass is chosen so that $m_0 x_0^{-3}$ matches the density of water. Due to a corresponding rescaling of the permittivity $\epsilon$ we can, without loss of generality, set the lattice unit of charge to the elementary charge $e_0=e.$\par
One limitation of the simple pseudopotential method we use is that the parameter $G$ controls both the interface width $\lambda_I$ and the surface tension $\gamma_{sc}.$ If we choose the time scale $t_0$ in such a way that the simulated fluid viscosity matches the viscosity of water, then for values of $\tau= 1$, which are preferable in single-relaxation-time BGK for reasons of accuracy, we get a time step of $t_0\approx 0.2$ps. At such a small time step the surface tension obtained e.g.~for $G=5.5x_0^3m_0^{-1}t_0^{-2}$ corresponds to 36 times the surface tension of water at room temperature. While one could lower $G$ to match the surface tension somewhat better, this would result in slower demixing dynamics, increased computational cost and extremely large interface widths of tens of nanometers. Instead, we choose $t_0$ in such a way, that we match the desired surface tension. As a side effect the viscosity in our simulations is significantly smaller than that of water, however we do not expect this to make any qualitative difference. The main source of fluid flow in our system is demixing, which is a diffusive process independent of inertial effects as long as $G$ is not chosen too large. While the resulting interface width is still of the order of a few nanometers, this is not neccessarily unrealistic for systems close to the critical point of demixing~\cite{buhn_molecular_2004,pousaneh_molecular_2016}, where structure formation by antagonistic salts is observed in experiments. When approaching the critical point, the interface length $\lambda_I$ diverges and the surface tension $\gamma_{sc}$ vanishes with a strong temperature sensitivity.\par
Choosing for example $x_0=$1nm, $t_0=10$ps and $G=5.5x_0^3m_0^{-1}t_0^{-2}$ we have $\gamma_{sc}=0.036m_0t_0^{-2}=2.75\cdot 10^{-4}$N~m$^{-1}$ and $\lambda_I=1.3x_0=1.3$nm. Accordingly, $\epsilon_r=40$ corresponds to $\epsilon=0.1$ in simulation units, $k_BT=0.6m_0x_0^2t_0^{-2}$ for a temperature of about 330K and a salt concentration of 0.1 is equivalent to about 166mmol l$^{-1}.$ In practice, as our aim is to understand the system fundamentally, we experimented with a wide range of simulation parameters without focusing very much on their relation to experimental values. For example in most simulations we keep $\Delta\mu$ in a range of about $3-8k_BT$, which is noticeably lower than the $15k_BT$ of NaBPh$_4$ commonly used in experiments, because this allows us to more easily avoid numerical errors from steep ion concentration gradients and high fluid forcing terms. Doing so we have come to the conclusion that, above all, the dimensionless numbers $\Lambda$ and $\lambda_d/\lambda_{ws}$ introduced in Eqs.~(\ref{eq:biglambda}) and (\ref{eq:fluxratio}) determine the behaviour of the system. In the interest of reproducibility, we nonetheless give the simulation parameters used in simulation units in all figure captions.
\section{Results}
\label{sec:results}
\subsection{Fluid-charge coupling}
\label{sec:coupling}
\begin{figure}
  \centering
  \includegraphics[width=0.49\textwidth]{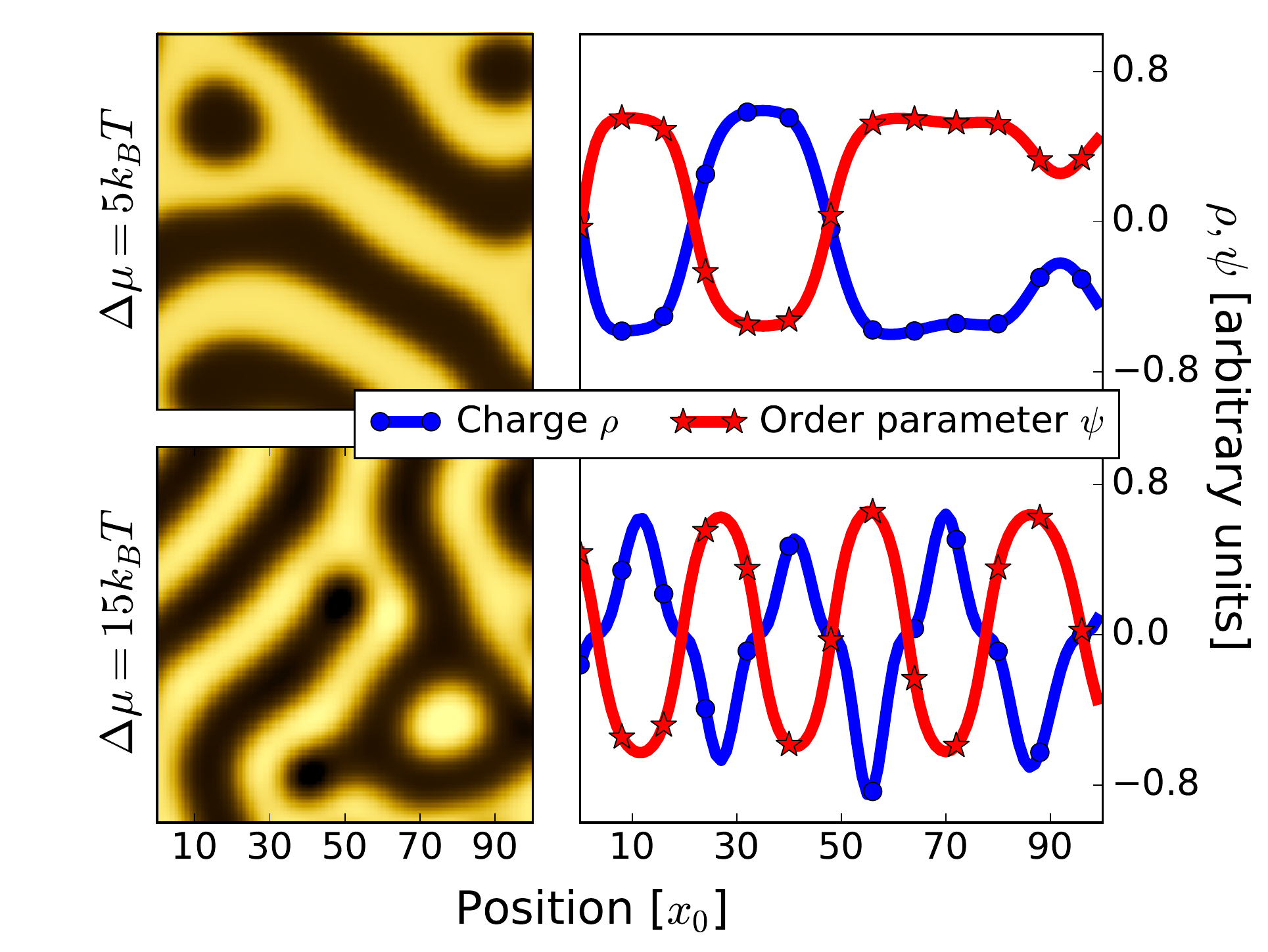}
  \caption{Left: Excerpts of the order parameter $\psi$ from 2D simulations. Right: Qualitative comparison of $\psi$ and charge $\rho$ in 1D cuts of length 100$x_0.$ $\rho$ is in arbitrary units in order to be on the same scale as $\psi.$ Top: $\Delta\mu=5k_BT.$ Bottom: $\Delta\mu=15k_BT.$ Parameters: $c_s=5\cdot10^{-5},$ $\epsilon=1.4\cdot 10^{-4},$ $k_BT=100,$ $G=4.5$}
  \label{fig:opcharge}
\end{figure}
\begin{figure*}
  \centering
  \includegraphics[width=0.424\textwidth]{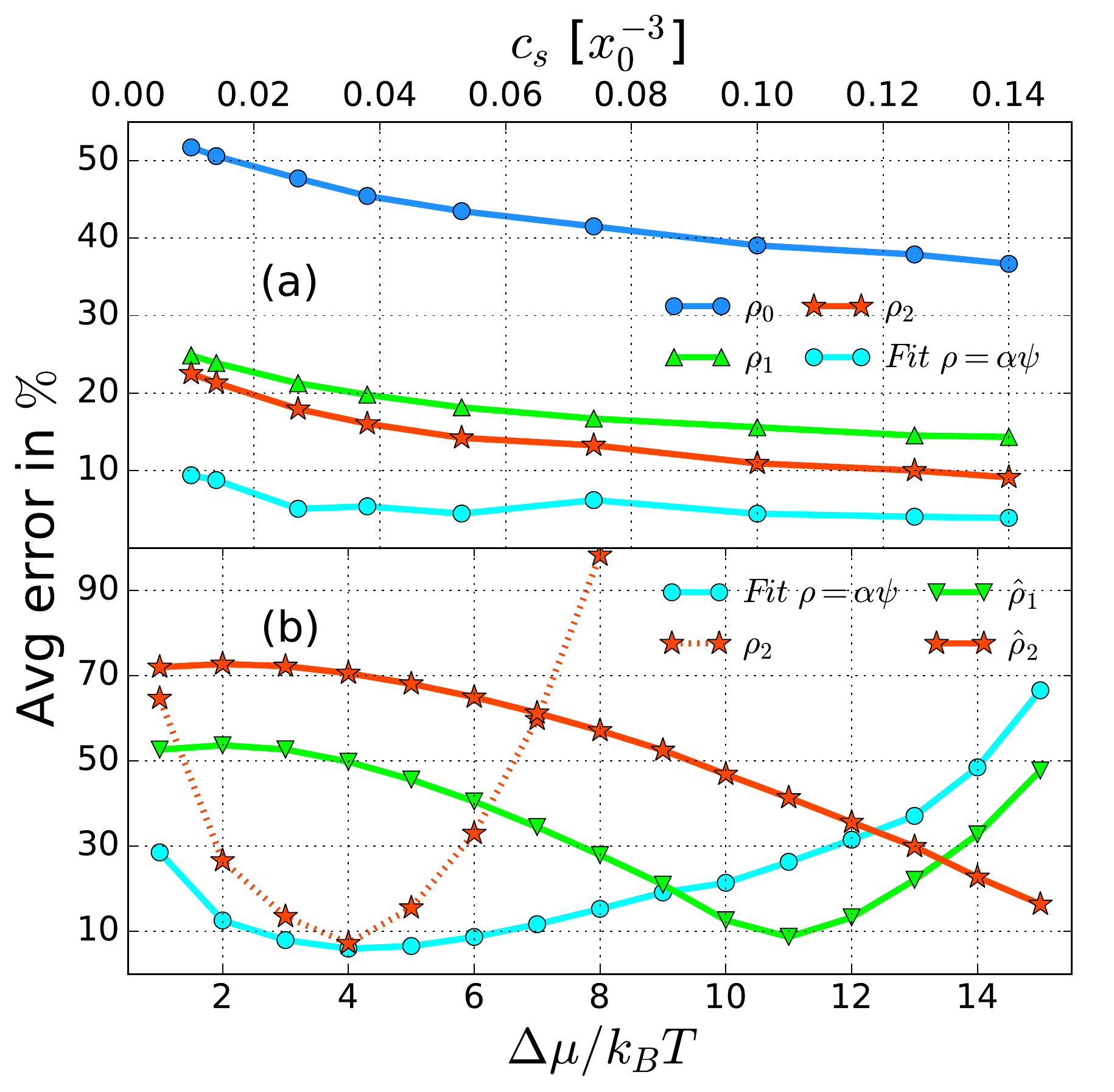}
  \includegraphics[width=0.57\textwidth]{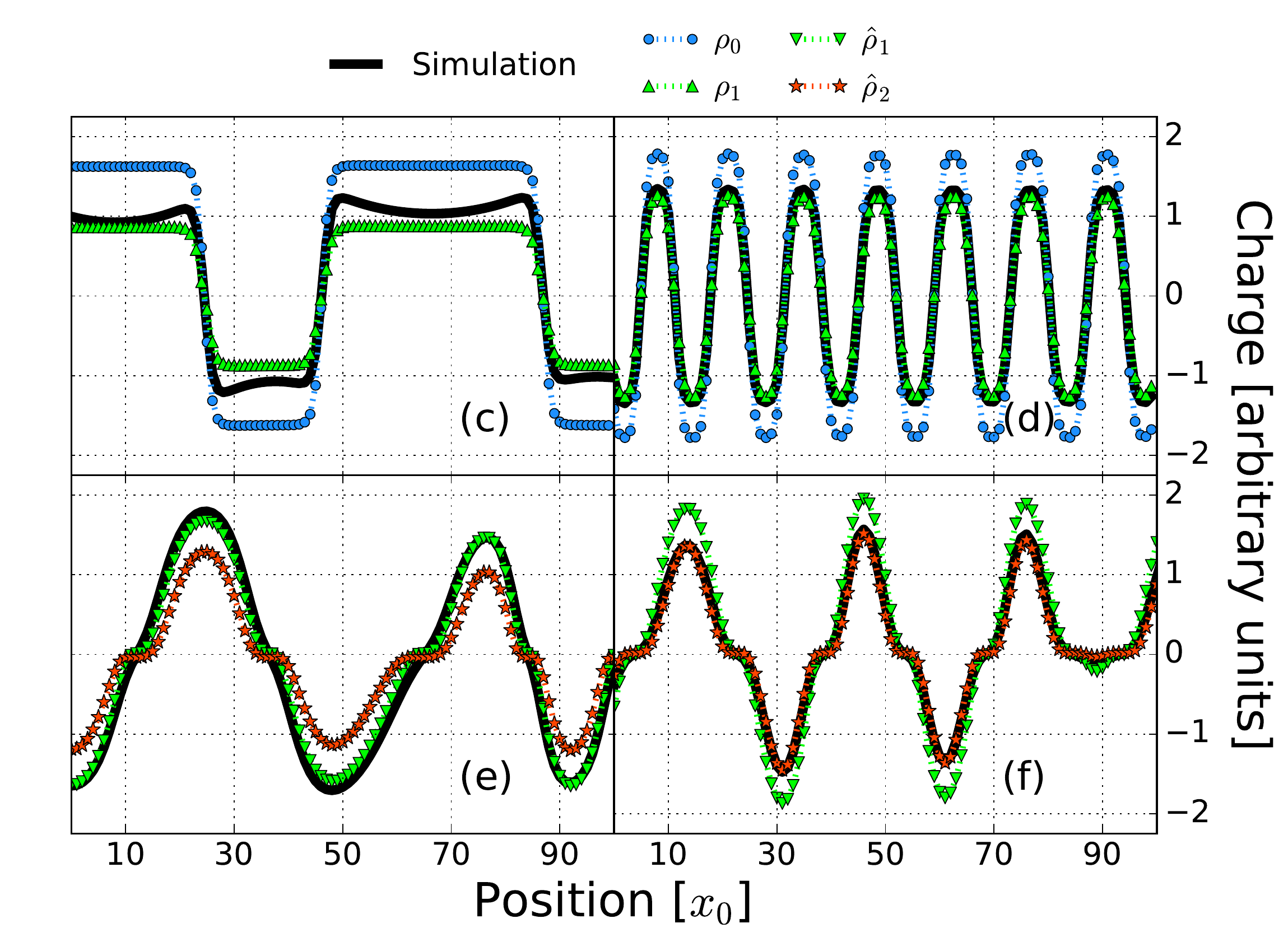}
  \caption{Errors in approximating the charge distribution via the order parameter while neglecting electrostatics. The charge approximations $\rho_n$ refer to the $n$-th order approximations introduced in Eq.~(\ref{eq:ion-expansions}). The hatted charge approximations $\hat{\rho}_n$ refer to the approximations with a normalizing prefactor accounting for finite salt concentrations introduced in Eqs.(\ref{eq:ionsum-eq-hidmu}) and (\ref{eq:charge-eq-hidmu}). Root mean square error $\sqrt{\langle (\rho-\rho_n)^2 \rangle/\langle \rho^2 \rangle}$ as a function of the salt concentration $c_s$ (a) and the antagonicity $\Delta\mu$ (b) for several orders of approximation $n$ as well as for a linear model $\rho=\alpha \psi$ using $\alpha$ as a fit parameter. Exemplary cuts through 2D charge distributions are shown for $c_s=0.01x_0^{-3}$ (c), $c_s=0.14x_0^{-3}$ (d), $\Delta\mu=10k_BT$ (e), and $\Delta\mu=15k_BT$ (f). Parameters: $\epsilon=4,$ $k_BT=1,$ $\Delta\mu=3k_BT,$ $G=5.5$ [(a), (c), and (d)], and $c_s=5\cdot10^{-5},$ $\epsilon=1.4\cdot 10^{-4},$ $k_BT=100,$ $G=4.5$ [(b), (e), and (f)].}
  \label{fig:rmserror}
\end{figure*}
All our simulations are initialized as a homogenous mixture with symmetric volume fraction $m=\langle n_1 \rangle/(\langle n_1 \rangle+\langle n_2 \rangle)=0.5$ and constant ion concentrations $c_\pm=0.5c_s$ everywhere given by the initial salt concentration $c_s.$ The fluid concentrations $n_1$ and $n_2$ are initially perturbed by a small stochastic prefactor at each lattice site in a way that keeps the overall volume fraction constant. For $G=5.5x_0^3m_0^{-1}t_0^{-2}$ spinodal decomposition during the first about one thousand time steps is driven almost completely by the pseudopotential forces of demixing. As the local contrast in fluid concentrations $n_1$ and $n_2$ grows, charges build up due to ion separation by solvation and the resulting electrostatic forces on the fluid arrest the process of coarsening. After a further about one thousand time steps we observe lamellar, bicontinuous structures.\par
Because of a strong coupling via the solvation potential, both the fluid order parameter $\psi$ and the charge $\rho$ follow the same lamellar pattern, as shown in Fig.~\ref{fig:opcharge}. Our key result in this paper is an analytic description of this coupling. Using the linear coupling limit valid for low $\Delta\mu/k_BT$ allows us to use prior results on the Ohta-Kawasaki free energy model to derive the two dimensionless numbers $\Lambda$ and $\lambda_d/\lambda_{ws}$ describing our system.\par
As we are concerned with systems close to the critical point of demixing, the pseudopotential forces of demixing are small and the time scale of fluid relaxation is much larger than that of ion relaxation. This leaves the advection term in Eq.~(\ref{eq:cahnhilliard}) largely irrelevant. In our simulations the ions are practically in equilibrium at all times with respect to the current fluid composition $\psi.$ From the equilibrium solution of the evolution Eq.~(\ref{eq:cahnhilliard}) for the charge $\rho$ and the total local ion concentration $c_t=c_++c_-$, we obtain two coupled differential equations,
\begin{align}
  \label{eq:omega-eq1}
  \nabla^2 \rho/e&=M\big( c_t \nabla^2 \Omega+\nabla c_t\cdot \nabla \Omega \big),\\
  \label{eq:omega-eq2}
  e\nabla^2 c_t&=M\big( \rho \nabla^2 \Omega+\nabla \rho\cdot \nabla \Omega \big).
\end{align}
For ease of notation we write $M=-\Delta\mu/\Delta\psi k_BT$ and $\Omega=\psi + e \phi \Delta\psi /\Delta\mu.$ When $M\lesssim 1$, solvation is weak compared to the ideal pressure of ions and diffusion keeps the total ion concentration homogenous, so that $c_t\approx c_0= \mathrm{const.}=c_s=\Gamma_0.$ Using $c_t=c_0$ our first approximation of the charge follows from Eq.~(\ref{eq:omega-eq1}) as $\rho_0/e=c_s M \Omega -\langle c_s M \Omega \rangle=c_s M\Omega+\Rho_0$, where $\Rho_0$ is chosen to ensure charge neutrality. Here, as throughout the rest of the paper, $\langle \cdot \rangle$ signifies a spatial average. $\rho_0$ is a valid approximation as long as $\rho$ does not depend linearly on spatial coordinates, which might be the case, e.g., if $\phi$ includes strong external electric fields. By inserting this into Eq.~(\ref{eq:omega-eq2}), we approximate the first order fluctuations of $c_t$ around its expectation value $c_s$, which will be significant for larger $M$:
\begin{equation}
\begin{aligned}
  \label{eq:ionsum-eq}
  e\nabla^2 c_1&=M\big(\rho_0\nabla^2\Omega+\nabla\rho_0\cdot\nabla\Omega \big)= \frac{M^2\Gamma_0}{2}\nabla^2 \Omega^2+M\Rho_0\nabla^2 \Omega\\
\rightarrow ec_1&=\frac{M^2\Gamma_0}{2}\Omega^2+M\Rho_0\Omega\underbrace{-\langle \frac{M^2\Gamma_0}{2}\Omega^2+M\Rho_0\Omega \rangle +ec_s}_{=\Gamma_1}.
\end{aligned}
\end{equation}
Here, $\Gamma_1$ is chosen to ensure total ion conservation. The approximation $c_t=c_1$ can in turn be plugged back into Eq.~(\ref{eq:omega-eq1}), yielding a higher-order approximation $\rho_1$ for the charge $\rho$, and so on. The general expansions of $\rho$ and $c_t$ up to order $n$ are
\begin{equation}
\begin{gathered}
  \label{eq:ion-expansions}
  \frac{\rho_n}{e}=\!\!\!\!\sum_{i=1,3,5,..}^{2n+1}\!\! \frac{M^i\Gamma_{n-(i-1)/2}}{i!}\Omega^i+\!\!\!\!\sum_{j=2,4,6,..}^{2n} \!\!\frac{M^j\Rho_{n-j/2}}{j!}\Omega^j+\Rho_n,\\
  e c_n=\!\!\!\!\sum_{i=1,3,5,..}^{2n-1}\!\! \frac{M^i\Rho_{n-(i-1)/2}}{i!}\Omega^i+\!\!\!\!\sum_{j=2,4,6,..}^{2n} \!\!\frac{M^j\Gamma_{n-j/2}}{j!}\Omega^j+\Gamma_n,
\\
\Rho_n=-\langle \rho_n/e-\Rho_n \rangle, \hspace{7.5ex} \Gamma_n=-\langle c_n-\Gamma_n \rangle +ec_s.
\end{gathered}
\end{equation}
It is worth noting that $\langle \Omega^k \rangle=0$ for uneven $k$ whenever the volume fraction is $1/2$ and the two fluids are treated identically apart from $\Delta \mu_+=-\Delta\mu_-$ and $z_+=-z_-.$ This in turn means that all $P_n=0$ and thus Eq.~(\ref{eq:ion-expansions}) becomes much simpler, with $\rho_n$ turning into a polynomial of only uneven powers of $\Omega$ and $c_n$ of only even powers of $\Omega.$\par
Eq.~(\ref{eq:ion-expansions}) fails by overestimating the fluctuations of $c_t$ when $\Delta \mu \gg k_BT.$ Strong solvation leads to completely ionless regions of $c_t\approx 0$ in the vicinity of the interfaces, because here both ion kinds experience solvation forces in opposite directions towards higher concentrations of their preferred fluid components. Numerically, whenever outgoing fluxes at any lattice site would reduce the local ion concentration below zero, all outgoing fluxes at that site are reduced by a common factor, so that the local ion concentration goes to zero instead. Ingoing fluxes are then calculated and potentially similarly reduced individually by the previously computed factor reducing the neighbouring lattice site's outgoing fluxes. In our simulations, ion concentrations may become negative without this type of discretization correction when $\Delta \mu \gtrsim 7k_BT.$\par
The analytic approximations leading to Eq.~(\ref{eq:ion-expansions}) do not take a limited ion availability into account and similarly predict negative $c_t$ for $\Delta \mu \gg k_BT.$ To a surprisingly good degree of accuracy we can still approximate $c_t$ in this limit by rescaling each consecutive $c_n$ by a factor $\Upsilon_n$ in order to force an amplitude of ion concentration fluctuations equal to $c_s.$ The additive constants $\Gamma_n$ then all become $0$ for $n\geq 1.$ We write the charge and ion concentration approximations including the correction for finite salt concentration as $\hat{\rho}_n$ and $\hat{c}_n.$ Thus Eq.~(\ref{eq:ionsum-eq}) becomes
\begin{equation}
\begin{gathered}
  \label{eq:ionsum-eq-hidmu}
\hat{c}_1=c_s\frac{\frac{M^2c_s}{2}\Omega^2+M\Rho_0\Omega}{\Upsilon_1},\\
\Upsilon_1=\bigg\langle \frac{M^2c_s}{2}\Omega^2+M\Rho_0\Omega \bigg\rangle,
\end{gathered}
\end{equation}
and by reinserting into Eq.~(\ref{eq:omega-eq1}) we obtain the charge as before.
\begin{equation}
% \begin{gathered}
  \label{eq:charge-eq-hidmu}
\hat{\rho}_1/e=\frac{M^3c_s^2}{24\Upsilon_1}\Omega^3+\frac{M^2\Rho_0c_s}{2\Upsilon_1}\Omega^2\underbrace{-\bigg\langle \frac{M^3c_s^2}{24\Upsilon_1}\Omega^3+\frac{M^2\Rho_0c_s}{2\Upsilon_1}\Omega^2 \bigg\rangle}_{=\Rho_1}.
% \end{gathered}
\end{equation}
Naturally, the procedure can be repeated to obtain higher order corrections and a general expansion as in Eq.~(\ref{eq:ion-expansions}). While $\rho_n$ and $c_n$ tend to converge towards the exact results $\rho$ and $c_t$ for $n\to \infty$ as long as $\Delta \mu$ is small enough that ion-depletion at the interfaces does not play a role (roughly $\Delta \mu < 5k_BT$), the ideal value of $n$ for $\hat{\rho}_n$ and $\hat{c}_n$ must be empirically chosen depending on $\Delta \mu$. We find that $n=2$ produces a more accurate approximation than $n=1$ only for high values of $\Delta \mu \approx 15k_BT$, the largest antagonicity we used in our simulations. Higher values of $n$ would likely only be appropriate for extreme antagonicities of $\Delta \mu\gg 15k_BT.$\par
The proposed methods of approximating the ion distributions gain their value from the possibility of neglecting electrostatic fluxes, so that $\Omega\approx \psi.$ In Fig.~\ref{fig:rmserror} we show the quality of the theoretical model using $\Omega=\psi$ for a range of values of $c_s$ and $\Delta\mu.$ In Fig.~\ref{fig:rmserror}a the root mean square errors of approximating $\rho$ as $\rho_n$ are seen to decrease with increasing $n$ and $c_s.$ In Fig.~\ref{fig:rmserror}c we can see that, for low $c_s$ and thus large structure sizes, the charge begins to concentrate at the interfaces due to electrostatic interaction with the counter charges in the opposite fluid component. This feature is not captured by $\rho_n$ due to neglecting electrostatics.\par
Errors depend in a more complicated manner on $\Delta \mu$ than on $c_s.$ As shown in Fig.~\ref{fig:rmserror}b, $\rho_2$, by ignoring limited availability of ions, quickly fails for high $\Delta \mu$ by predicting negative ion concentrations. For $\Delta \mu\to 0$ it fails again because structure sizes diverge when there is no solvation, causing $\rho$ and $\psi$ to become decoupled. The charge approximations $\hat{\rho}_n$ corrected for limited ion availability fare much better for high $\Delta\mu$, but the errors here do not converge to 0 for $n\to \infty.$ Instead, $\hat{\rho}_1$ is a better approximation for intermediate $\Delta\mu$ and $\hat{\rho}_2$ for high $\Delta \mu.$ In Figs.~\ref{fig:rmserror}e and f we can see how for larger $n$, $\hat{\rho}_n$ features more pronounced regions of zero charge at the interfaces, which correspond to ion-depleted regions of $c_t\approx 0.$
\subsection{Ohta-Kawasaki equivalency}
\label{sec:ohta}
Using the analytical results for the charge and ion concentrations from Sec.~\ref{sec:coupling} we can significantly simplify our system and make use of known results from the Ohta-Kawasaki free energy model. Looking back at the solvation force in Eq.~(\ref{eq:fluidforce}) we find, by taking the linear approximation for symmetric fluids $\rho=\rho_0=e c_sM \psi$
\begin{equation}
  \label{eq:solvationforce}
  \vec{F}_s=-\frac{\Delta\mu}{\Delta\psi}\nabla \Big((c_+-c_-) \psi \Big)=-2c_sM \frac{\Delta\mu}{\Delta \psi} \psi \nabla \psi.
\end{equation}
This is very similar to the continuum limit of the pseudopotential force acting on fluid $\sigma$ due to interaction with fluid $\bar{\sigma}$ in the pseudopotential model when neglecting third and higher order gradient terms~\cite{sbragaglia_continuum_2009}:
\begin{equation}
  \label{eq:shanchenforce}
  \vec{F}^\sigma_{sc}=-G \Psi_\sigma \nabla \Psi_{\bar{\sigma}}.
\end{equation}
Setting $\Psi_\sigma=n_\sigma$ and postulating $\nabla n_2=-\nabla n_1$ due to incompressibility, we find that $\vec{F}_s$ in fact points in precisely the same direction as $\vec{F}^1_{sc}+\vec{F}^2_{sc}$, that is, away from the interface.
\begin{equation}
  \label{eq:dirsolvationforce}
\begin{split}
  \psi\nabla \psi= 2 \psi \frac{n_2\nabla n_1-n_1\nabla n_2}{(n_1+n_2)^2}=\\ = -2(n_1+n_2)^{-3}\Big(n_2 \nabla n_1 + n_1 \nabla n_2 \Big).
\end{split}
\end{equation}
This means that the solvation force acts purely to increase surface tension. When fluid incompressibility is fulfilled, $n_1+n_2$ is constant and Eq.~(\ref{eq:dirsolvationforce}) has the same shape as Eq.~(\ref{eq:shanchenforce}). Similarly, the ideal pressure force resulting from ion diffusion also acts only in the direction of the pseudopotential force assuming symmetric fluids and therewith $c_t=c_1=\frac{M^2\Gamma_0}{2e} \psi^2+\Gamma_1/e.$
\begin{equation}
  \label{eq:dirdiffforce}
  \vec{F}_d=-k_BT\nabla c_t=-k_BT M^2\Gamma_0 \psi \nabla \psi/e.
\end{equation}
Indeed we find both $\vec{F}_d$ and $\vec{F}_s$ to be almost irrelevant in determining the general morphology of the system in simulations as long as their magnitude does not rise to a level where they cause fluid compression. This holds true even when $\vec{F}_d$ and $\vec{F}_s$ are orders of magnitude larger than the electrostatic force $\vec{F}_e.$ By subsuming the influence of $\vec{F}_d$ and $\vec{F}_s$ into a modified surface tension $\gamma=\gamma_{sc}+\gamma_d+\gamma_s$ and neglecting the electrostatic fluxes $\vec{j}_\pm^e$, we are left with a greatly simplified system. The surface tension $\gamma$ can be determined from Eq.~(\ref{eq:eqstate}), e.g.~via a Laplace test, meaning that a single bubble of one fluid component of various radii is initialized in the bulk of the other fluid component. According to the Young-Laplace law, the surface tension $\gamma$ is the slope of the graph of pressure differences inside to outside the droplet versus the inverse droplet radius.\par
Whenever $\Delta\mu$ is small enough for the first-order charge approximation $\rho_0/e=c_sM\psi$ to be accurate, our system is fully equivalent to the Ohta-Kawasaki model commonly used to model diblock copolymers~\cite{spadaro_uniform_2009}:
\begin{equation}
  \label{eq:ohta}
  \begin{split}
  \mathcal{F}_T=\mathcal{F}_F(\psi)+\int \xi |\nabla \Phi |^2 \hspace{1ex} \mathrm{d\mathbf{r}},\\
  - \nabla^2 \Phi=\psi.
  \end{split}
\end{equation}
The first term is the fluid demixing energy $\mathcal{F}_F$ introduced in section \ref{sec:method} but with $\kappa_{sc}$ substituted with $\kappa$, which fulfills $\gamma=0.5 \int_{-\infty}^{+\infty}\kappa|\nabla \psi|^2\, \mathrm{d}x$ for a hyperbolic tangent shaped $\psi$ and thus includes the surface tension contributions of $\vec{F}_d$ and $\vec{F}_s$. Crucially the model adds an electrostatic term weighted by $\xi$ in which the order parameter directly corresponds to a charge. The long-range interaction potential $\Phi$ is, in the first-order charge approximation $\rho=\rho_0$, proportional to the electrostatic potential with $\Phi=\phi \epsilon/ec_sM.$\par
The Ohta-Kawasaki model is usually treated in either of two limits depending on the characteristic length $\lambda_L$ of structures in equilibrium as compared to the interface width $\lambda_I.$ In the weak-segregation limit, i.e.~when $\lambda_L\lesssim \lambda_I$, the wavelength of structures is~\cite{weith_stability_2013} $\lambda_{ws}=2\pi (\kappa/2\xi)^{1/4}$, or, as $\kappa = 6\gamma\lambda_I/\Delta\psi^2$, $\lambda_{ws} \propto (\gamma/\xi)^{\frac{1}{4}}.$ In the strong-segregation limit~\cite{weith_stability_2013}, i.e.~when $\lambda_L\gg \lambda_I$, $\lambda_{ss}\propto (\gamma/\xi)^{\frac{1}{3}}.$ In analogy to our free energy Eq.~(\ref{eq:iongibbs}), the term $\xi |\nabla \Phi |^2$ corresponds to the electrostatic term $0.5 \rho\phi.$ By setting $\rho=\rho_0=e c_sM \psi$ we can see by comparison that $\xi=(\Delta\mu/k_BT\Delta\psi)^2e^2c_s^2/2\epsilon=\Lambda\gamma/2\Delta\psi^2 \lambda_I^3.$ In summary:
\begin{equation}
\begin{gathered}
  \label{eq:biglambda}
  \Lambda=\frac{\lambda_I^3}{\epsilon\gamma}\bigg(\frac{ec_s\Delta\mu}{k_BT} \bigg)^2,\\
   \lambda_{ws}= 2\pi\lambda_I \bigg( \frac{6}{\Lambda} \bigg)^{\frac{1}{4}}=2\pi\sqrt{\frac{k_BT}{e c_s \Delta \mu}} \big( 6\lambda_I\epsilon\gamma \big)^{\frac{1}{4}},\\
   \lambda_{ss}= 4 \lambda_I \bigg(\frac{12}{ \Lambda} \bigg)^{\frac{1}{3}}=4\bigg(\frac{k_BT}{e c_s \Delta\mu} \bigg)^{\frac{2}{3}} \big(12\epsilon\gamma \big)^{\frac{1}{3}}.
\end{gathered}
\end{equation}
The dimensionless number $\Lambda$ turns out to be useful to predict the morphology produced by a given set of system parameters. In our simulations we find by varying all involved simulation parameters, including the pseudopotential interaction parameter $G$, that structures begin to gradually dissolve in the range of $\Lambda\approx 1-10$, or $\lambda_{ws}/\lambda_I\approx 6-9$ up to a return to a mixed state at higher $\Lambda.$ The onset of structure dissolution is characterized by lamellar regions interspersed with blots of mixed regions.\par
Following Onuki and Okamoto~\cite{onuki_solvation_2009} the electrostatic surface tension contribution in a system with heterogenity in only the $z$ direction, such as parallel lamellar structures, can be calculated by subtracting the electrostatic energy density integrated from one bulk to the other over an interface:
\begin{equation}
  \label{eq:surftens}
  \gamma_{\rm{eff}}=\gamma - \int \xi|\nabla \Phi|^2 \mathrm{d}z=\gamma\bigg( 1-\frac{\Lambda}{2\lambda_I^3}\int \frac{|\nabla\Phi|^2}{\Delta\psi^2}\mathrm{d}z \bigg).
\end{equation}
By making a simple 1D ansatz of $\psi=0.5\Delta\psi\, \mathrm{sin}(2\pi x/\lambda_{ws})$ valid in the weak-segregation limit and integrating $|\nabla \Phi|^2$ from the bulk of one phase to the other over $\lambda_{ws}/2$, we can predict a critical $\Lambda=2^{13}  3^{-3} \pi^{-4} \approx 3.1$, at which $\gamma_{\rm_{eff}}$ is expected to become zero. This approximately matches the point at which we observe the onset of structure dissolution in simulations.\par
Recall that in deriving the equivalency to the Ohta-Kawasaki model we set $\Omega\approx \psi$, i.e.~we neglected electrostatic fluxes. Defining the interfacial energy density $W_i=0.5\kappa|\nabla\psi|^2$ and the electrostatic energy density $W_e=0.5\epsilon |\nabla \phi|^2$ we can, using again $\rho=\rho_0$, quantify the ratio of solvation and electrostatic fluxes as
\begin{equation}
  \label{eq:fluxratio}
  \frac{|\vec{j}_\pm^s|}{|\vec{j}_\pm^e|}=4\pi^2\frac{\lambda_d^2}{\lambda_{ws}^2}\sqrt{\frac{W_i}{W_e}}=\frac{\Delta\mu}{\sqrt{6\lambda_I\gamma e^2/\epsilon}}\sqrt{\frac{W_i}{W_e}}.
\end{equation}
Here we introduce the Debye length $\lambda_d=\sqrt{\epsilon k_BT/e^2c_s}.$ It gives the length scale over which the density of countercharges near a surface charge decays. The interfacial and electrostatic energy densities are the main two counteracting factors driving structure formation in the Ohta-Kawasaki model, with the former striving to minimize interface area and the latter favouring either homogenous charge, or rapid spatial oscillations of $\rho$ and therewith $\psi$, leading to a large number of interfaces. A balance of these two energy densities is found when lamellar structures form, so it seems reasonable to assume a comparable order of magnitude in such morphologies. Indeed, Araki and Onuki showed this to be true in steady-states and weak segregation~\cite{araki_dynamics_2009}. Assuming $W_i\approx W_e$, Eq.~(\ref{eq:fluxratio}) suggests electrostatic fluxes to be about an order of magnitude smaller than solvation fluxes and hence quantitatively negligible whenever the Debye length is $\gtrsim \lambda_{ws}/2.$ In other words, for $\lambda_d\gtrsim \lambda_{ws}/2$ ion dynamics are largely unaffected by electrostatic interactions with other ions. When $\lambda_d\to \lambda_{ws}/2$ structure sizes diverge and for $\lambda_d<\lambda_{ws}/2$ spinodal decomposition is largely unaffected by electrostatic effects.
\subsection{Nematic ordering}
\label{sec:nematic}
\begin{figure*}
  \centering
  \includegraphics[width=0.8\textwidth]{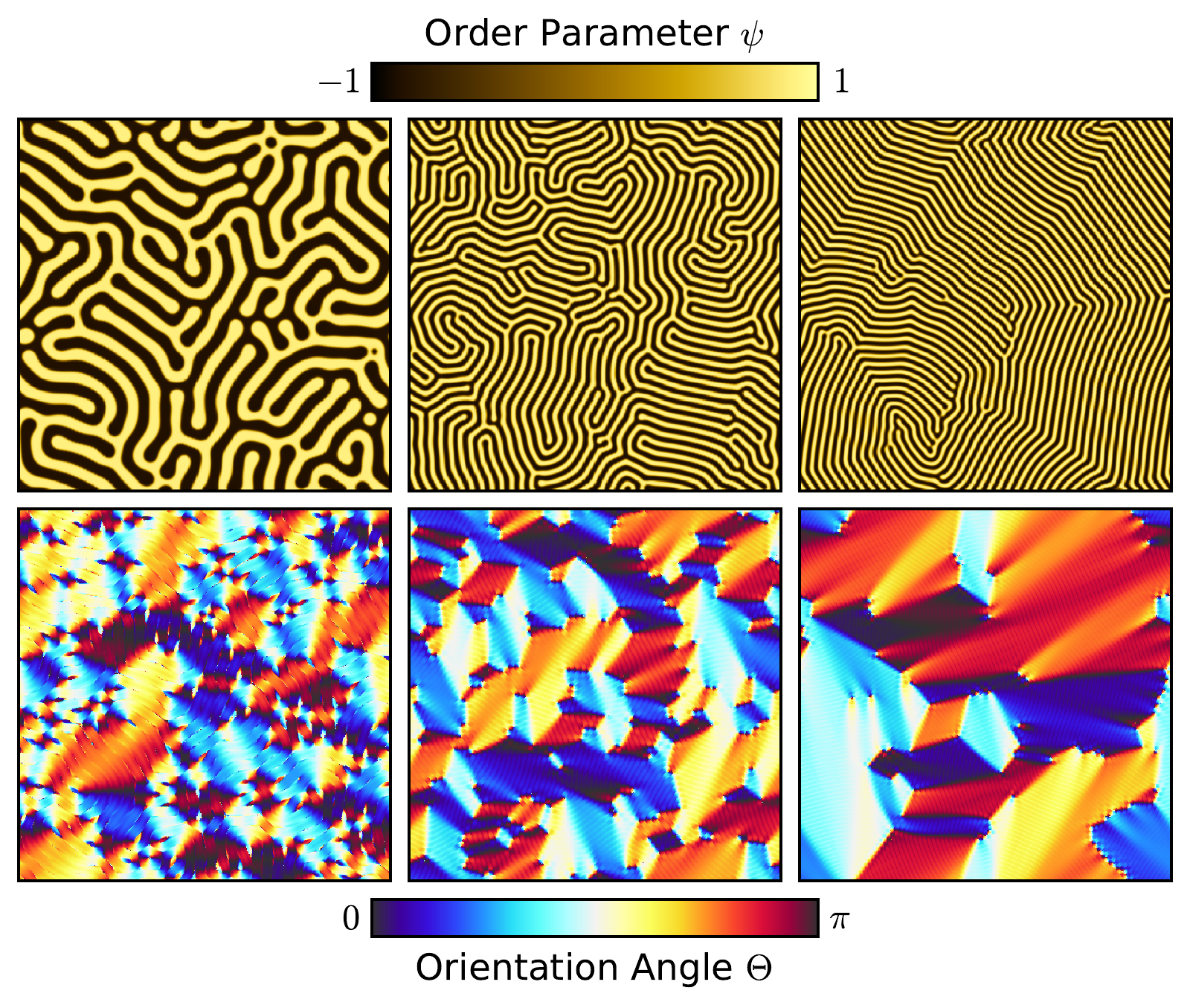}
  \caption{From left to right, $\Lambda$ increases from 0.08 $(c_s=0.027x_0^{-3})$, to 0.9 $(c_s=0.1x_0^{-3})$, and finally 1.3 $(c_s=0.14x_0^{-3}).$ Top: Order parameter $\psi$ after 750000 time steps in a 504x504 system. Bottom: Local value of the lamellar alignment angle $\Theta$ defined in Eq.~(\ref{eq:theta}). The angle is measured clockwise starting from vertical lamellae. Blue represents lamellae oriented from the bottom left to the top right (or vice versa), red from the top left to the bottom right. Parameters: $\epsilon=4,$ $k_BT=1,$ $\Delta\mu=3k_BT,$ $G=5.5.$}
  \label{fig:ordercombo}
\end{figure*}
\begin{figure}
  \centering
  \includegraphics[width=0.49\textwidth]{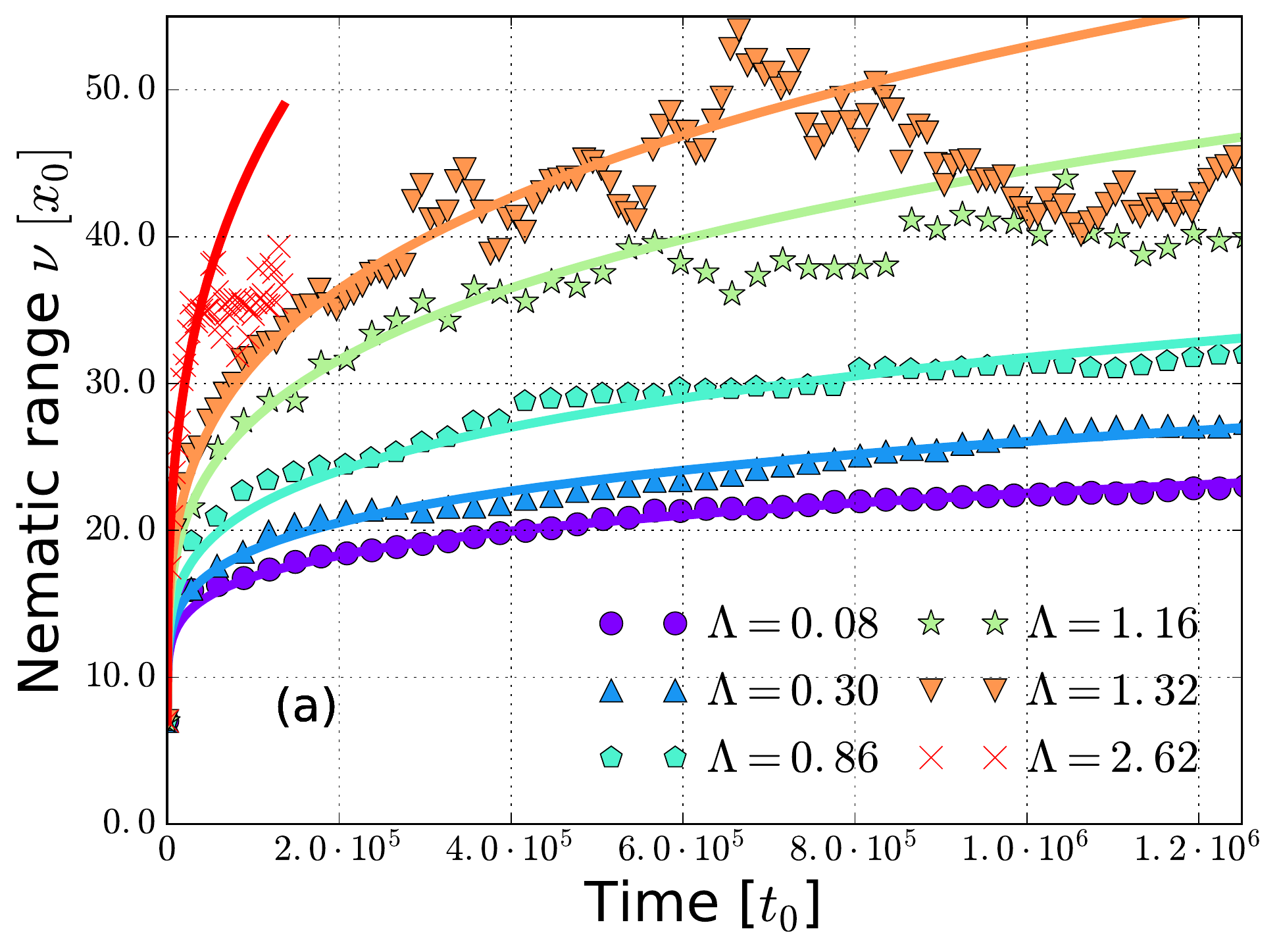}
  \includegraphics[width=0.235\textwidth]{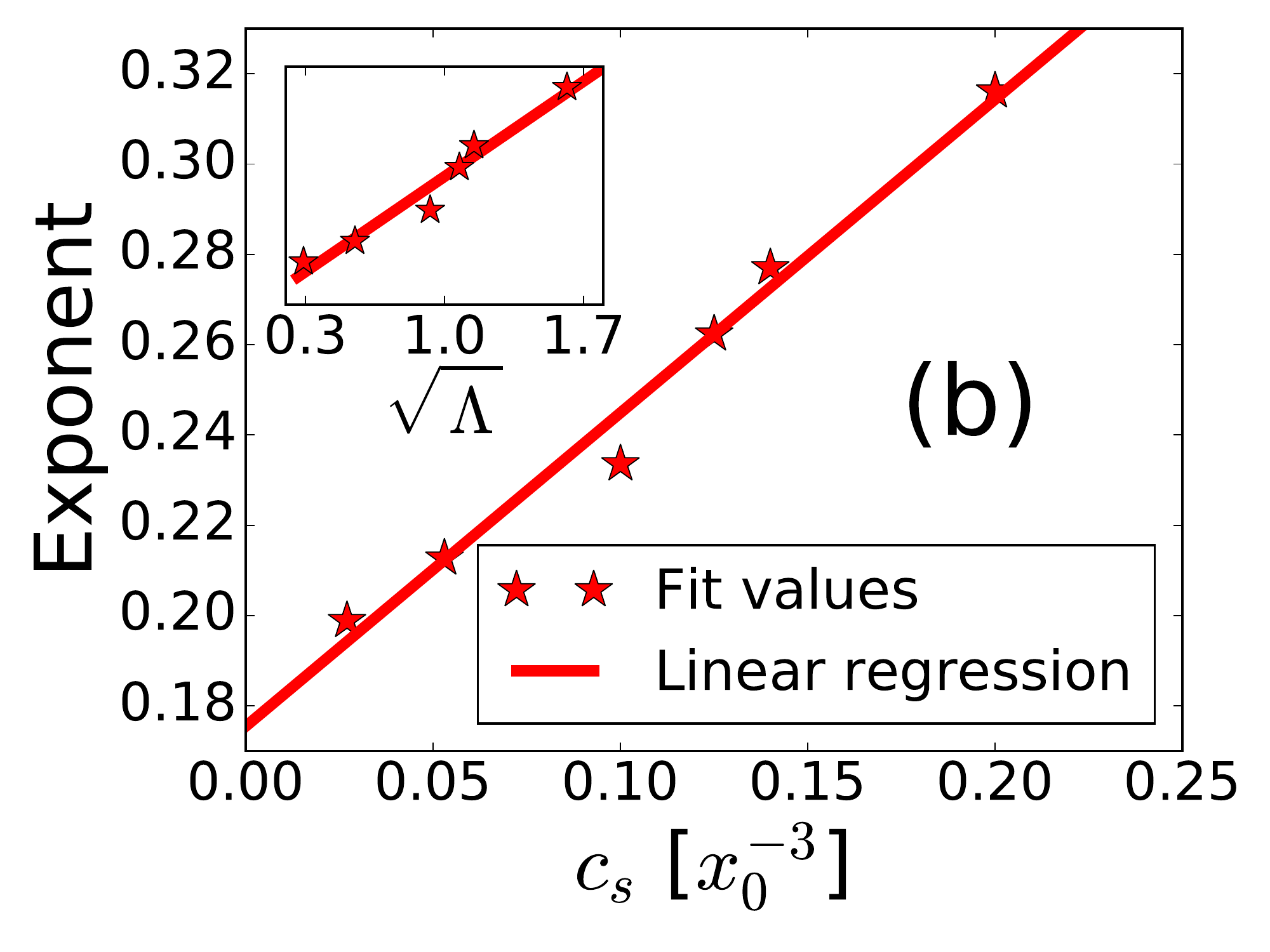}
  \includegraphics[width=0.235\textwidth]{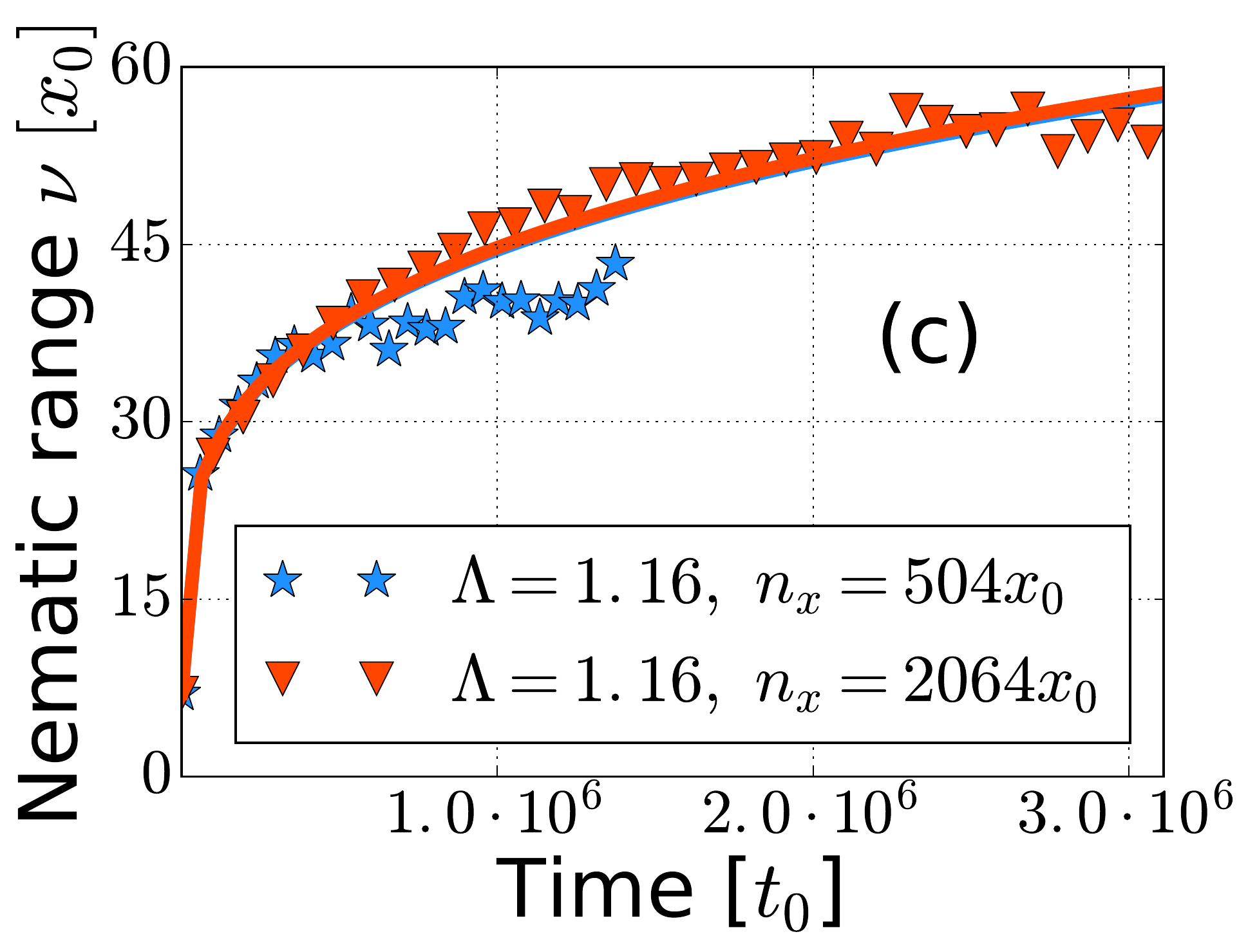}
  \caption{(a): Range of nematic ordering over time for five different salt concentrations. Ordering tends to increase for all salt concentrations with time, but at a much slower rate for lower salt concentrations. $\nu$ fluctuates strongly at high nematic range $\nu\gtrsim 38x_0$ due to finite size effects. (b): The exponent of the power law fit above as a function of the salt concentration and as a function of $\sqrt{\Lambda}$ (inlet). The results are slightly different because of the salt concentration dependence of $\gamma.$ The exponent is determined using only data for $\nu< 38x_0.$ (c): Time evolution of $\nu$ in a larger 2064$x_0$ size system compared to the results above in a 504$x_0$ system. The fluctuations in $\nu$ visible for $\nu\gtrsim 38x_0$ do not appear in larger systems and can be attributed to finite size effects. The power law fit (full lines) over the full data set in the 2064$x_0$ system is practically identical to that of the 504$x_0$ system for $\nu<38x_0$. Parameters: $\epsilon=4,$ $k_BT=1,$ $\Delta\mu=3k_BT,$ $G=5.5.$}
  \label{fig:nematic}
\end{figure}As long as the ion dynamics are dominated by solvation, the morphology is almost completely determined by $\Lambda.$ As $\Lambda$ is increased, lamellae become thinner and the degree of nematic ordering at a given time increases. In the top row of Fig.~\ref{fig:ordercombo} we show three representative morphologies for $\Lambda=0.08$, 0.9 and 1.3.\par
Nematic order parameters commonly used in the analysis of liquid crystal structures measure the overall homogenity of nematic orientation. They are suitable when one is interested in deviations from some prefered direction, for example given by an applied electric field, surface patterning or initially ordered state~\cite{buluy_optical_2018,luders_order_2015,saintillan_orientational_2007}, or to determine the nematic homogenity of some spontaneously ordered steady-state~\cite{purdy_measuring_2003,ginelli_large-scale_2010}. In our case, ordering progresses without any inherently preferred direction and we are interested in the time evolution of the size of nematically ordered domains. As long as these domains are much smaller than the simulation domain, standard nematic order parameters indicate no nematic order, because all nematic orientations are equally frequent in the simulation domain. When the size of nematically ordered domains approaches the simulation domain size, some random nematic orientation begins to dominate and standard nematic order parameters indicate order, but we are uninterested in this regime because under these conditions finite size effects affect the time evolution of nematic order.\par
Instead we quantify nematic ordering over time by a nematic range parameter $\nu$ giving the length scale over which the nematic orientation changes. For this purpose we first determine the angle of orientation $\Theta$ of the lamellae at a given position. The bottom row of Fig.~\ref{fig:ordercombo} shows the local angle of orientation as a function of position for three morphologies. The gradient magnitude of the orientation field $\Theta$ gives the average angle by which the lamellar orientation changes over a distance of one lattice site in the direction of the greatest rate of change. The local nematic range $\nu$ estimates the distance in lattice sites in direction of fastest change of $\Theta$ over which $\Theta$ changes to a perpendicular orientation. Using the fact that the gradient of $\psi$ ought to be everywhere perpendicular to the local lamella, we choose the definition
\begin{equation}
  \label{eq:theta}
\begin{gathered}
  \Theta=\mathrm{arctan}\,\bigg(\frac{\partial_x \psi}{\partial_y \psi}\bigg)+\frac{\pi}{2}\\
  \nu=\frac{\pi}{2} \frac{1}{\langle |\nabla \Theta | \rangle}.
\end{gathered}
\end{equation}
When the entire system domain is filled with exactly parallel lamellae, $\nabla \Theta$ is zero everywhere, and $\nu \to \infty.$ Otherwise, the smaller the domains of common orientation are, the larger $|\nabla \Theta|$ is on average, so that $\nu \to 0$ for disordered systems.\par
The driving force of nematic ordering is long-range electrostatic fluid forcing. As our model does not include thermal fluctuations, no opposing force exists to disrupt nematic ordering. Hence, nematic ordering appears to continue indefinitely, until either the entire system domain is filled by perfectly ordered lamellae, or the process of nematic ordering is hindered by finite size effects. In Figs.~\ref{fig:nematic}a and c, we show how $\nu$ increases mostly monotonously in time. Nematic ordering begins to fluctuate considerably when domains of common orientation become comparable in size to the simulation domain, which occurs in Fig.~\ref{fig:nematic}a when $\nu\gtrsim 38x_0.$ While this corresponds to only about 8\% of the system size of $504x_0$, recall that $\nu$ measures the size of ordered domains in the direction of the fastest change of $\Theta.$ In any other direction, the domain may be considerably larger. One example of finite size effects contributing to fluctuations of $\nu$ is the orientation angle dependence of the number of separate lamellae that fit into the simulation domain due to its square shape and the use of periodic boundary conditions. For this reason large numbers of nematic defects must form, leading to decreasing $\nu$, before nematic ordering can increase further.\par
As nematic ordering is driven by electrostatic fluid forcing, and the electrostatic fluid forcing is proportional in magnitude to $\Lambda$, it is not surprising that ordering progresses faster for higher $\Lambda.$ Nematic range as a function of time seems to be well-modelled as a power law of the form $\nu/x_0=(t/t_0)^\eta$, with an exponent $\eta\propto c_s$ proportional to the salt concentration. Because of the early onset of finite size effects we made another simulation on a much larger 2064$x_0$ system over $3\cdot 10^6$ time steps for $\Lambda=1.16$. The resulting evolution of $\nu$ shown in Fig.~\ref{fig:nematic}c confirms the previously mentioned power law behaviour but without any major deviations due to finite size effects. Such a power law behaviour has been previously described for the growth of nematically ordered domains in the Ohta-Kawasaki model~\cite{huang_coarse-grained_2007}. As shown in Eq.~\ref{eq:biglambda}, $\Lambda \propto c_s^2.$ Judging from smaller-scale simulations, the system parameters $\epsilon$ and $\Delta \mu$ affecting the magnitude of electrostatic forces and therewith $\Lambda$ appear to similarly affect the rate of nematic ordering. We suspect that indeed $\eta \propto \sqrt{\Lambda}$, though we have so far only performed sufficiently long-term simulations to accurately determine $\eta$ as a function of the salt concentration. The results of fitting the exponents $\eta$ as either $\propto c_s$ or $\propto \sqrt{\Lambda}$ are shown in Fig.~\ref{fig:nematic}b. We are currently engaged in implementing a performance efficient version of our code for 2D simulations using the D2Q9 scheme in order to further study nematic ordering as a function of $\Lambda$, for larger system sizes and time scales, and to test the influence of viscosity. It is known that in the Ohta-Kawasaki model, $\eta$ is a function of $\xi\propto \Lambda$, though the precise form of this dependency has not been studied to our knowledge~\cite{huang_coarse-grained_2007}.\par
Because the lamellar patterns are oppositely charged, the nematic ordering can be significantly sped up and oriented in a controlled manner by applying an external electric field. The electric field has to be applied early during the demixing process to be able to align the lamellae. When an electric field is applied on a domain of already nematically ordered lamellae, the resulting electrostatic forces acting on both sides of an interface are directed in opposition to each other and cancel out due to opposite charges. Increasing the field strength eventually leads to a break-up of the lamellar structures which is known as the Helfrich-Hurault instability in the context of smectic and cholesteric liquid crystals~\cite{onuki_electric_1995}. In the case of our simulations, hydrodynamic chaos ensues, though in the Stokes regime, a square lattice undulating pattern can be predicted from free energy considerations~\cite{onuki_electric_1995,fukuda_dynamics_1995}. If the field strength is decreased again, lamellar structures in alignment with the external field direction will reform. Structure formation by electric fields has been extensively studied for diblock-copolymer mixtures~\cite{boker_large_2002,kan_tuning_2016,pester_block_2017} with results that should be entirely transferable to the case of antagonistic salt mixtures as long as external electric fields are not strong enough to decouple the ions from the fluid order parameter.
\subsection{Length scales}
\label{sec:lengths}
\begin{figure}
  \centering
  \includegraphics[width=0.45\textwidth]{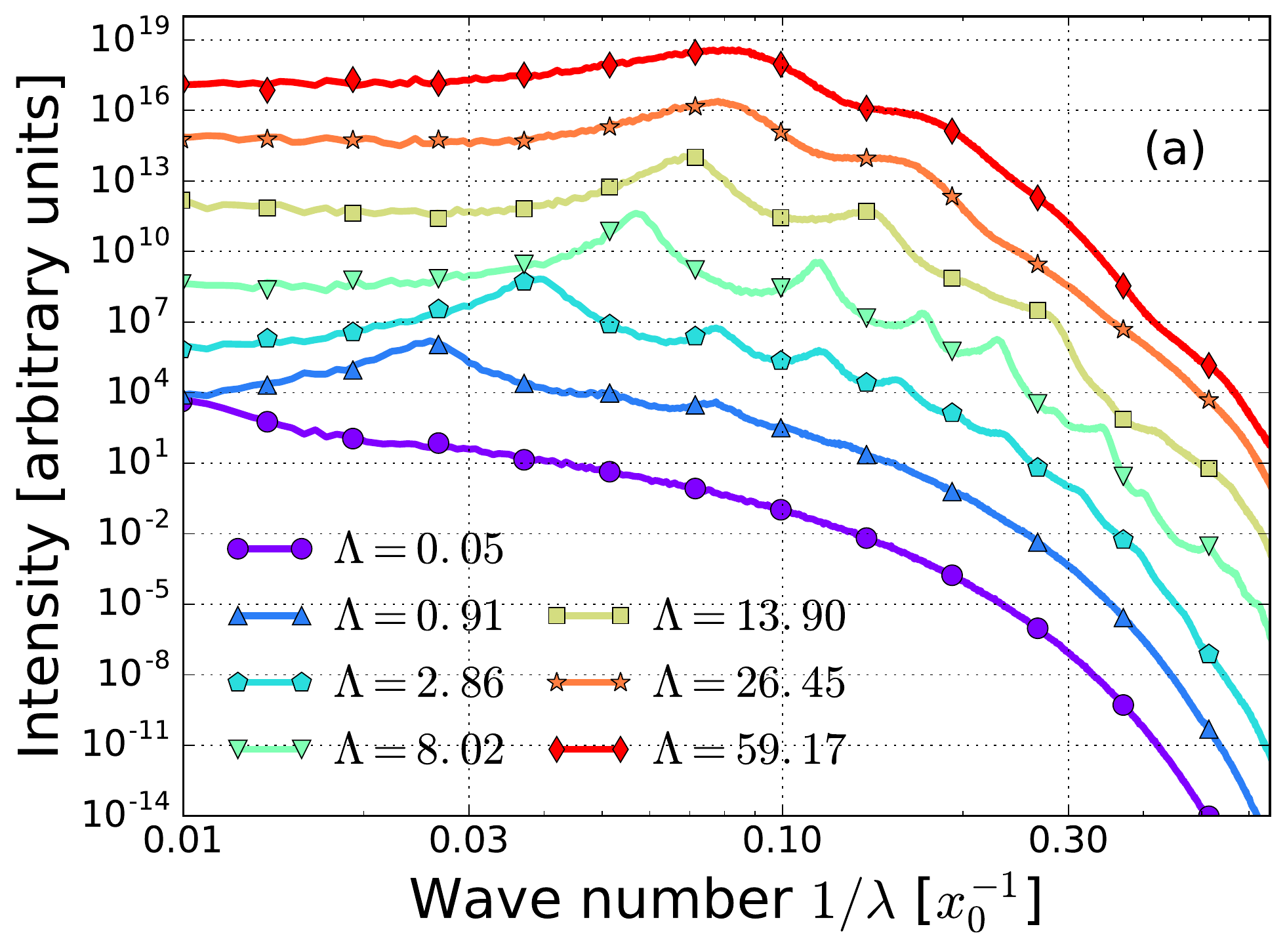}
  \includegraphics[width=0.4\textwidth]{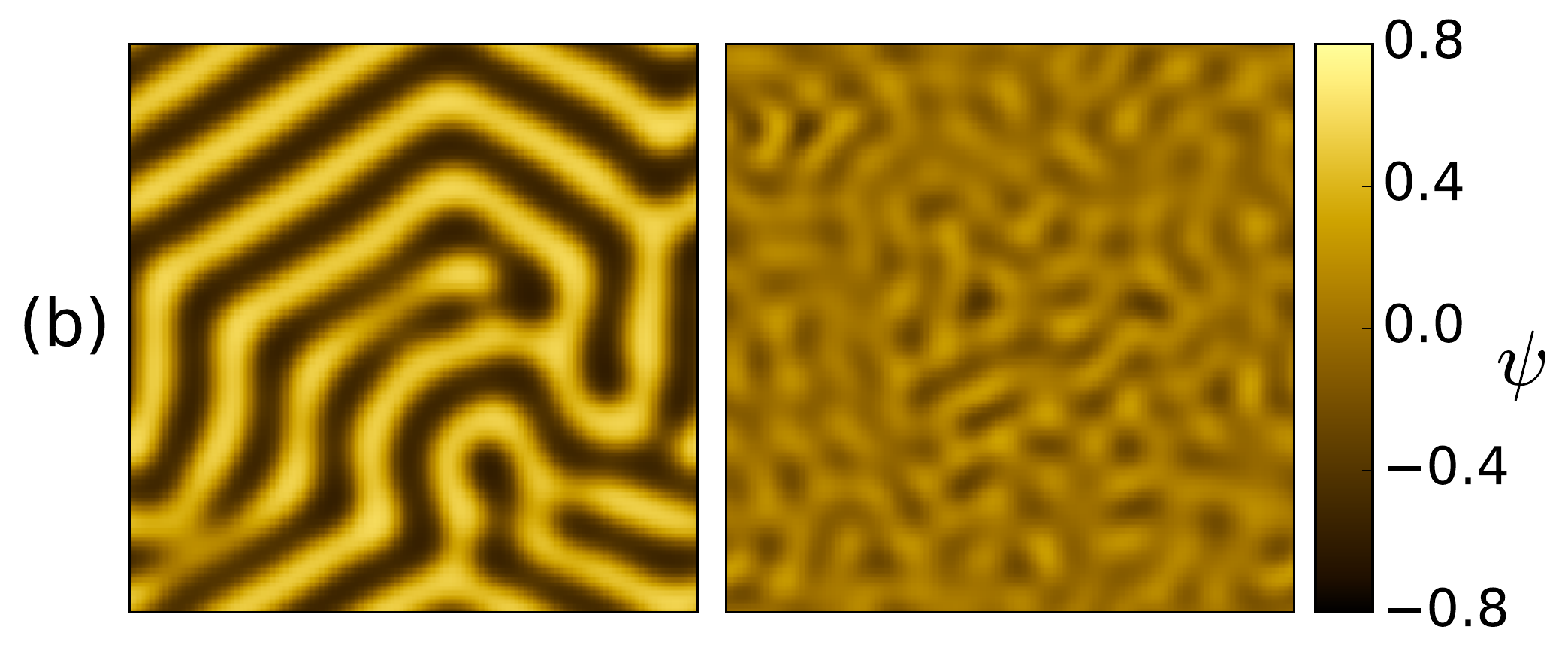}
  \caption{(a): Simulation results of power spectral density showing primary and secondary structure peaks and disappearance of structure at high $\Lambda.$ $\Lambda$ was varied by changing salt concentration. For readability each graph after $\Lambda=0.05$ is shifted by a factor of $10^3$ along the y-axis compared to the previous one. (b): Order parameter on identical color scales for $c_s=2\cdot 10^{-7}x_0^{-3}$ ($\Lambda\approx 8$) on the left and five times that salt concentration ($\Lambda\approx 59$) on the right. Parameters: $c_s=(1-100)\cdot 10^{-8},$ $\epsilon=2\cdot 10^{-10},$ $k_BT=10^5,$ $\Delta\mu=5k_BT,$ $G=4.5.$}
  \label{fig:simulationpsd}
\end{figure}In small-angle neutron scattering (SANS) a neutron diffraction image of a sample is recorded and time-averaged on a screen. According to the Rayleigh-Gans equation, this diffraction image is well-approximated as the radially averaged power spectral density (PSD) of the scattering length density in the sample. The scattering length density is an empirically determined material constant. Fig.~\ref{fig:simulationpsd}a shows the PSD $S_{n_s}(q)$, i.e.~the absolute squared of the Fourier transform, of a linear combination $n_s=s_1n_1+s_2n_2+s_+c_++s_-c_-$ of the fluid and ion concentrations. We choose the prefactors $s_\sigma$ and $s_\pm$ representing relative scattering length densities as $s_1=1.4$, $s_2=6.4$, $s_-=2.1$ and $s_+=0$ from the experimental values for a D$_2$O/3MP mixture with the antagonistic salt NaBPH$_4$~\cite{sadakane_membrane_2013}. Note that we do not aim to quantitatively match the experimental data, as we are operating at different volume fractions and neglecting potentially important system parameters such as permittivity and viscosity differences between the fluids. Nonetheless, $S_{n_s}(q)$ has a notable resemblance to experimental results from SANS~\cite{sadakane_membrane_2013}. For low salt concentrations, or low $\Lambda$, the PSD approximately follows the Ornstein-Zernike function~\cite{noauthor_accidental_1914,sadakane_membrane_2013}, indicating typical hydrodynamic concentration fluctuations but no periodic structure. At intermediate salt concentrations, a major peak appears at a wavenumber $q_m=1/\lambda_L$ corresponding to the center-to-center spacing $\lambda_L$ between two neighbouring lamellae of either fluid.\par
The secondary peaks at higher wavenumbers for intermediate $\Lambda$ in Fig.~\ref{fig:simulationpsd}a are the harmonics of the peak frequency, i.e.~they are located at integer multiples of $q_m.$ The uneven harmonics can be explained by approximating the lamellar structures as a square wave of frequency $q_m$, which by the Fourier series can be decomposed into the sum of all uneven harmonics of $q_m.$ The even harmonics are caused by a decrease of the total fluid density at the interfaces due to the pseudopotential forces of demixing. Our general assumption of fluid incompressibility is of course only approximately true here. The density dip at the interface can also be approximated as a step function, but at twice the frequency $q_m$ of the lamellae, as each lamellar has two interfaces. At least one such secondary peak at quite precisely double the frequency of the first peak is visible in the experimental data of Sadakane et al.~\cite{sadakane_membrane_2013}, and as its intensity is low compared to the noise, we believe that higher order peaks may have simply not been resolved due to imprecisions of the measurement. The PSD of the order parameter $S_\psi(q)$, while otherwise almost identical to $S_{n_s}(q)$, does not show secondary peaks at the even harmonics of $q_m.$ The sharp interfacial density dips in $n_1$ and $n_2$ are smoothed by normalization with $n_1+n_2.$\par
Going to high salt concentrations, we again recover essentially the same picture in the PSD as for low salt concentrations. As we illustrate in Fig.~\ref{fig:simulationpsd}b, periodic structures destabilize and remixing occurs, when $\Lambda$ exceeds some threshold. In section \ref{sec:ohta} we estimated the critical value where the effective surface tension goes to zero as $\Lambda_c\approx 3.1.$ In simulations the point where we observe structure dissolution varies depending on the values of $G$ and $\Delta \mu/k_BT$ in the range of about $\Lambda_c \approx 1-10.$ A likely reason for this is that the sinusoidal single-wave approximation we made in estimating $\Lambda_c$ is overly simplistic. The actual shape of the lamellae can have aspects of a square wave as well as flat shoulders of almost constant zero-valued order parameter at the interfaces for high $\Delta \mu$, as shown in Figs.~\ref{fig:rmserror}e and f. Also, an imperfect nematic ordering changes the electrostatic field in a very non-trivial way.\par
It is worth noting that the remixing state shown on the right in Fig.~\ref{fig:simulationpsd}b does not ever reach fluid flow equilibrium, with domains of non-zero order parameter sprouting and dissolving continuously. We expect that this likely unphysical behaviour might disappear when using a less coarse-grained model, where each fluid molecule experiences only the electrostatic force acting on ions it is currently bound to by solvation. As it is, electrostatic forces are applied equally to both fluid components at each lattice site.\par
\begin{figure*}
  \centering
\begin{tabular}{c c}
  \includegraphics[width=0.45\textwidth]{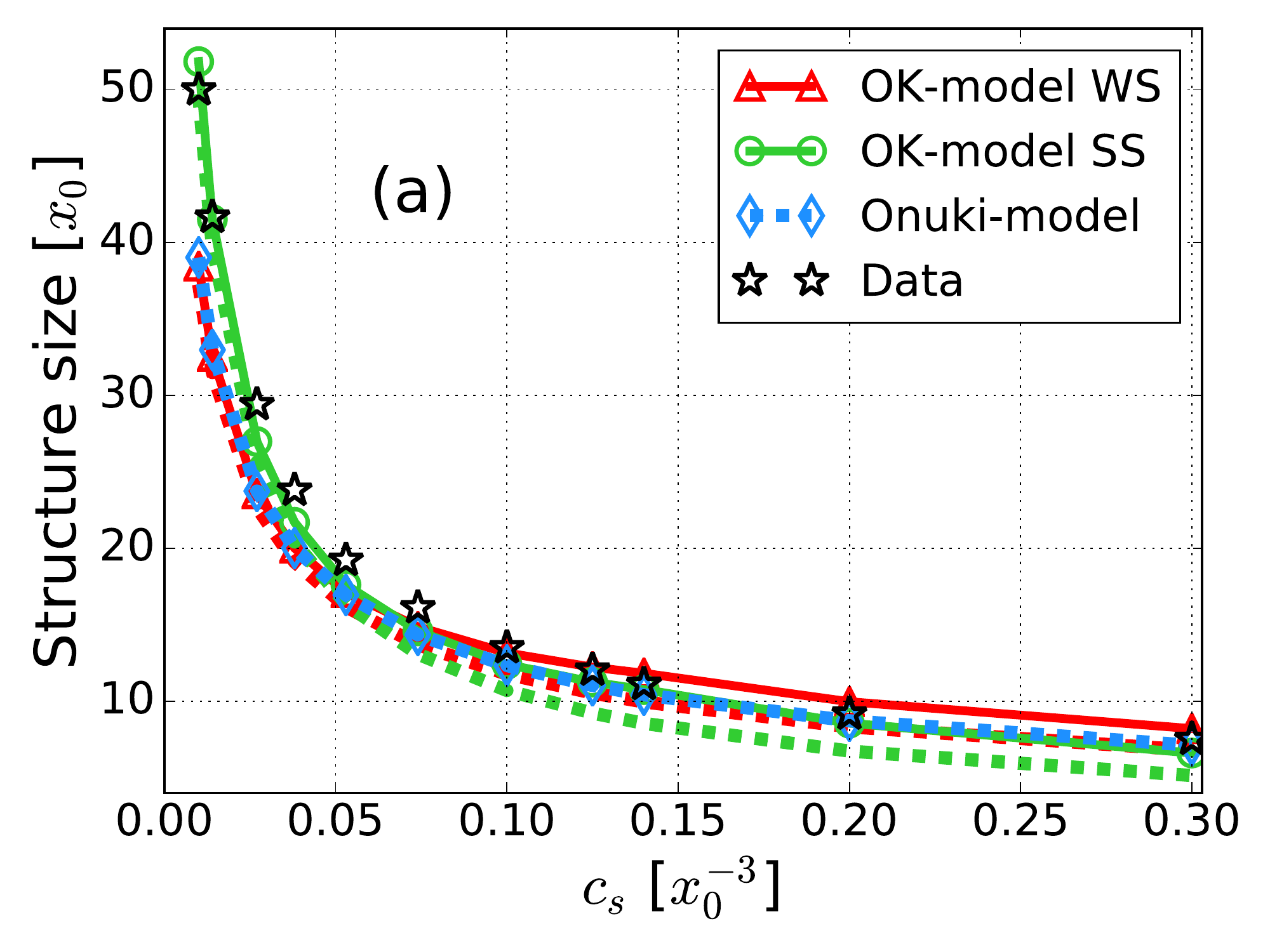}&
  \includegraphics[width=0.45\textwidth]{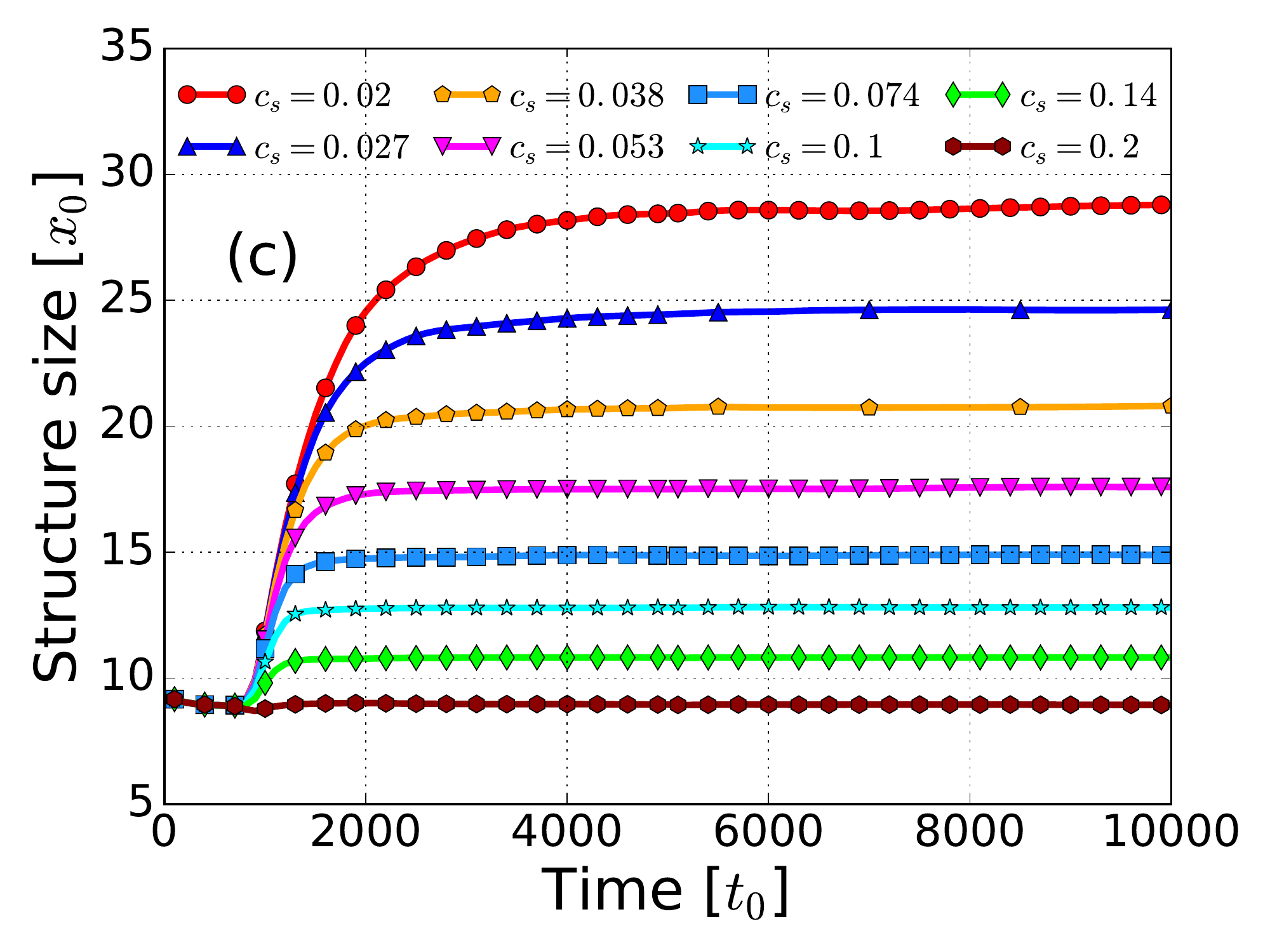}\\
  \includegraphics[width=0.45\textwidth]{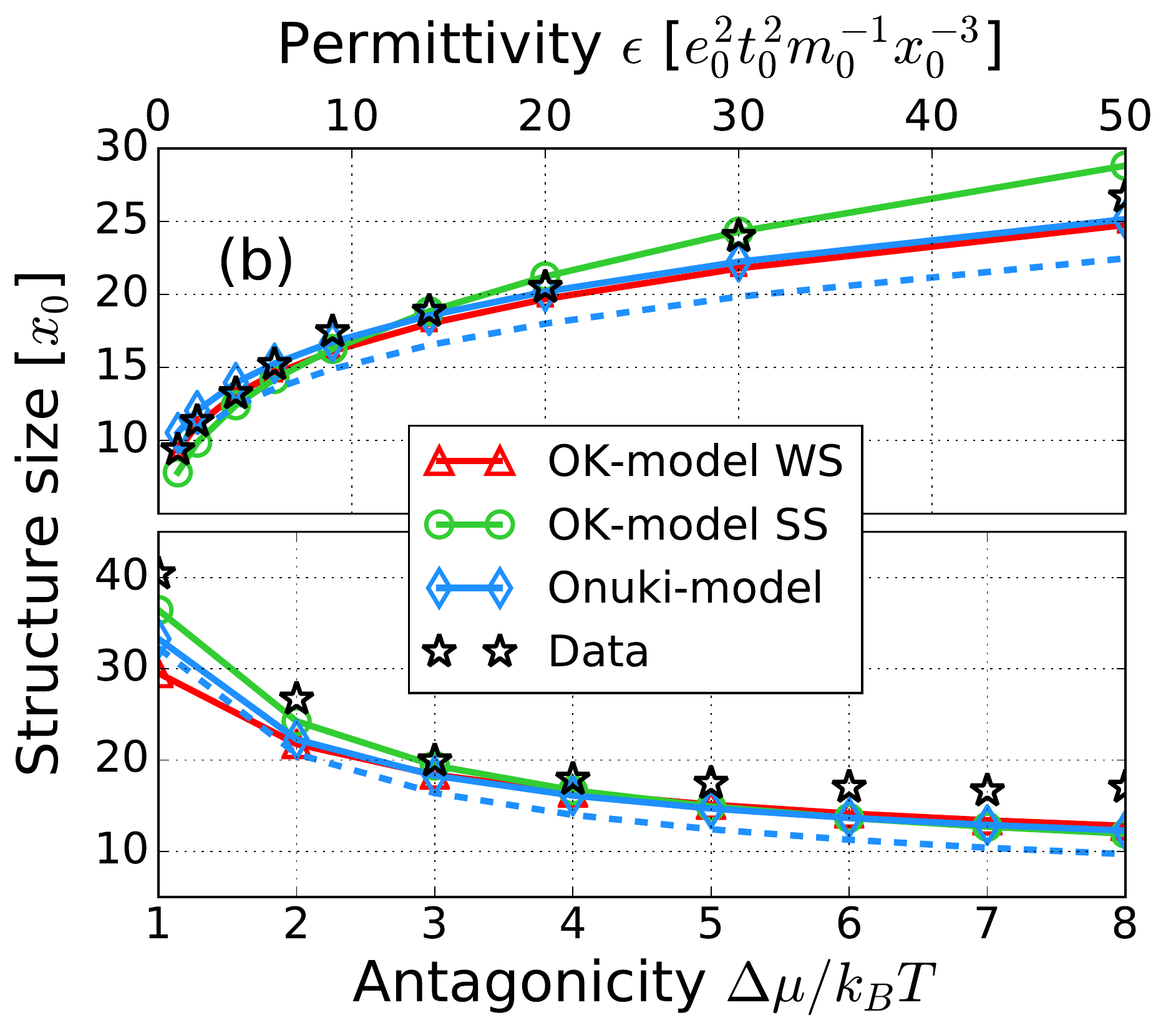}&
\begin{minipage}{0.4\textwidth}
\vspace{-7cm}
\caption{(a,b): Comparison of theoretical predictions in weak-segregation (WS) and strong-segregation (SS) limits (see Eq.~(\ref{eq:biglambda})), as well as of the theory of Onuki (see Eq.~(\ref{eq:onukiscale})) with simulation results of equilibrium structure sizes for (a): various salt concentrations and (b): for a range of permittivities and antagonicities. When varying $\epsilon$, we kept $c_s=0.1x_0^{-3}$, when varying $\Delta\mu$ $c_s=0.05x_0^{-3}.$ Full lines show the theoretical predictions using $\gamma$ including pressure contributions from solvation and the ideal pressure according to Eq.~(\ref{eq:eqstate}), while dashed lines result from neglecting all ionic contributions when calculating surface tension. (c): Time convergence of length scales for various salt concentrations. Parameters: $\epsilon=4,$ $k_BT=1,$ $\Delta\mu=3k_BT,$ $G=5.5.$}
\label{fig:lengthfig}
\end{minipage}
\end{tabular}
\end{figure*}
Extracting the structure size, i.e.~the periodicity $\lambda_L$ of the lamellae as the inverse of the spatial frequency of the primary peak in the PSD we find good agreement with the scaling laws derived in Eq.~(\ref{eq:biglambda}) according to the Ohta-Kawasaki model. We compare the theoretical predictions with simulation results for varying salt concentrations, dielectric permittivity and antagonicity in Figs.~\ref{fig:lengthfig}a and b. The parameters $\lambda_I$ and $\gamma$ are determined via Laplace tests. $\lambda_I$ is determined by fitting a hyperbolic tangent function to a 1D cut through the droplet interface. Full lines show the predicted structure sizes using a value of $\gamma$ obtained from Laplace tests including ionic contributions. Here the surface tension is calculated using Eq.~(\ref{eq:eqstate}), so that ion pressure and solvation effects are present but electrostatic contributions are disabled by setting $\vec{j}^e_\pm=\vec{F}_e=0$ in the Laplace test. For dashed lines the Laplace test is performed without any ions present, so that $\gamma=\gamma_{sc}.$ Although $\gamma$ changes by about a factor of 2 due to ionic contributions from the lowest to the highest salt concentration, the impact on the structure size predictions is not very large. Non-electrostatic ionic contributions to surface tension may be neglected in calculating the structure size unless a particularly high degree of accuracy is desired.\par
Following Onuki and Kitamura~\cite{onuki_solvation_2004}, the structure size is given by
\begin{equation}
  \begin{gathered}
  \label{eq:onukiscale}
  \lambda_o=\frac{2\pi \lambda_d}{\sqrt{\gamma_p-1}}=\sqrt{\frac{4\pi^2\epsilon k_BT}{e^2c_s(\gamma_p-1)}},\\
  \gamma_p=\frac{|\frac{\Delta\mu_+}{k_BT\Delta\psi}-\frac{\Delta\mu_-}{k_BT\Delta\psi}|}{2\sqrt{\kappa_{sc} e^2/\epsilon (k_BT)^2}}=\frac{\Delta\mu}{\sqrt{6\gamma_{sc} \lambda_I e^2/\epsilon}},
  \end{gathered}
\end{equation}
with a dimensionless parameter $\gamma_p$ quantifying the strength of antagonicity. Our weak-segregation scaling in Eq.~(\ref{eq:biglambda}) can be rewritten as $\lambda_{ws}=2\pi\lambda_d/\sqrt{\gamma_p}$, which is almost identical to Onuki's prediction. The model by Onuki predicts a divergence of structure sizes when $\gamma_p\to 1$, which our model does not reproduce directly. The reason for this is that the Ohta-Kawasaki model, from which the scaling laws in Eq.~(\ref{eq:biglambda}) are derived, describes our system accurately only when electrostatic fluxes are small versus solvation fluxes and thus $\gamma_p\gg 1$ (cf.~Eq.~(\ref{eq:fluxratio})). When $\gamma_p\lesssim 1$, electrostatic fluxes keep the ions bound to the interfaces, the charge is no longer strongly coupled to the order parameter and periodic structure formation ceases. We find Onuki's prediction to be almost identical to the weak-segregation Ohta-Kawasaki limit in Figs.~\ref{fig:lengthfig}a and b for $\gamma_p \approx 11.$ Onuki's model fares slightly worse than the strong-segregation Ohta-Kawasaki limit when structure sizes grow larger than about $15\lambda_I,$ but Onuki's model can be expected to fare better when $\gamma_p\to 1.$\par
The position of the primary spatial frequency peak, i.e.~the structure size, can also be quite accurately determined by calculating the radially averaged PSD of the order parameter $S_\psi(q)$, i.e.~the structure factor, and then taking its first moment:
\begin{equation}
  \label{eq:psdpeak}
  q_m=\frac{\sum_q S_\psi(q) q}{\sum_q S_\psi(q)}.
\end{equation}
The dominant length scale taken as the inverse of $q_m$ converges very quickly compared to the slow convergence of nematic ordering, as seen in Fig.~\ref{fig:lengthfig}c. Note the slower convergence for lower salt concentrations, as it is the electrostatic forcing, scaling with the salt concentration, that stops the spinodal decomposition in the first place. As structure sizes are increased at a constant interface width $\lambda_I$ we gradually approach the limit of strong segregation, in which Araki and Onuki similarly observed a noticeably slower convergence of structure sizes~\cite{araki_dynamics_2009}.\par
Comparing the structure factor of the order parameter in the model by Onuki and Kitamura~\cite{onuki_solvation_2004} with that in the model of Ohta and Kawasaki~\cite{ohta_equilibrium_1986} reveals why a divergence of structure sizes occurs in the former as $\gamma_p\to 1$ but not in the latter. In the Ohta-Kawasaki model the inverse structure factor can be approximated as
  \begin{equation}
    \label{eq:structure_ohta}
    S_\psi^{-1}(q)\propto q^2+Cq^{-2},
  \end{equation}
with some constant $C>0$ and a $q^{-2}$ term stemming from Coulombic attraction of opposite fluid phases inhibiting macroscopic demixing and forcing $S_\psi\to 0$ for $q\to 0$. In the description of pure diblock copolymer solutions this is of course a reasonable condition indicating that oppositely charged blocks of a single diblock copolymer cannot stretch and separate indefinitely~\cite{ohta_equilibrium_1986,schmid_theory_2010}. In the model of Onuki and Kitamura on the other hand~\cite{onuki_solvation_2004}
\begin{equation}
  \label{eq:structure_onuki}
  S_\psi^{-1}(q)\propto q^2+\gamma_p^2\frac{\lambda_d^{-4}}{q^2+\lambda_d^{-2}}.
\end{equation}
Due to the presence of the $\lambda_d^{-2}$ term, we no longer neccessarily have $S_\psi \to 0$ for $q\to 0$. The fact that long-range interactions of charges are screened by the Debye layer allows for macroscopic demixing. By performing the first and second derivatives of $S_\psi^{-1}(q)|_{q=0}$ we find that the structure factor $S_\psi(q)$ has a maximum at $q=0$ for $\gamma_p\leq 1$ and a minimum at $q=0$ for $\gamma_p>1$. Thus macroscopic demixing occurs for $\gamma_p\leq 1$. A similar form of structure factor as in the model of Onuki and Kitamura can also be used to describe polyelectrolytes stabilized by electrostatic repulsion in a poor solvent. In such systems mesophase structures are formed for small salt concentrations and a transition to macroscopic demixing occurs when $\lambda_d$ falls below some threshold value~\cite{joanny_weakly_1990,raphael_annealed_1990}.
\subsection{Extension to 3D}
\label{sec:3D}
\begin{figure*}
  \centering
  \includegraphics[width=0.9\textwidth]{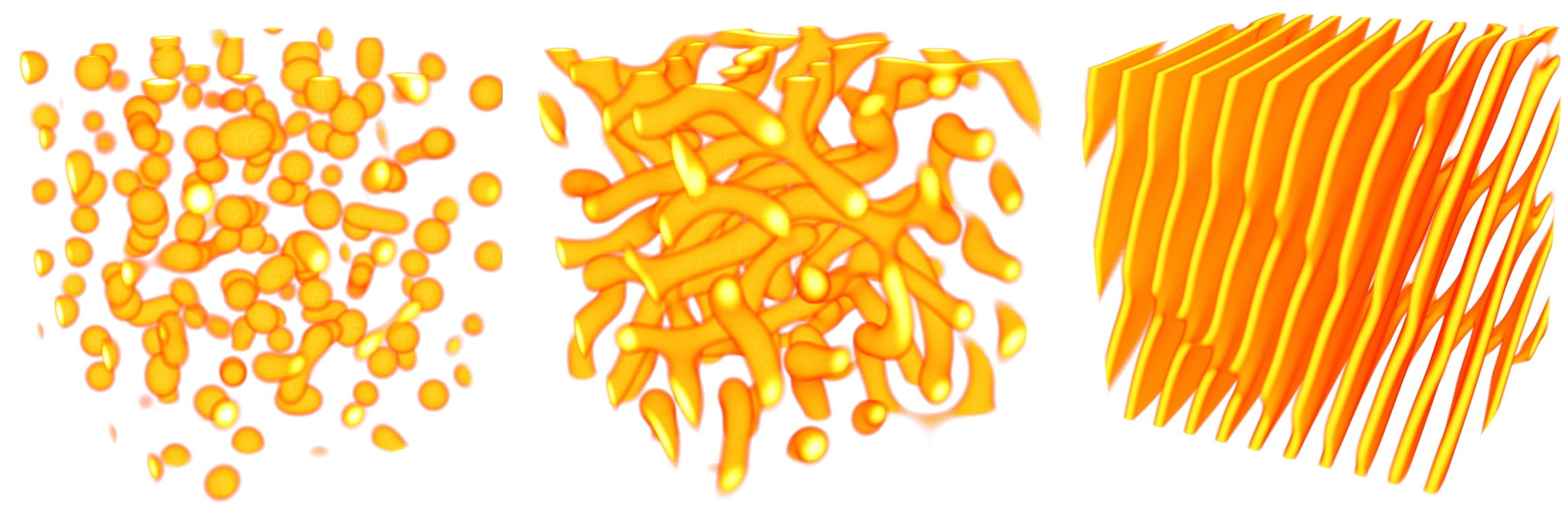}
  \caption{Resulting droplet, tubular and lamellar morphologies in 3D systems for $m=0.2,0.3$ and $0.5$ from left to right. Parameters: $c_s=0.125,$ $\epsilon=4,$ $k_BT=1,$ $\Delta\mu=3k_BT,$ $G=5.5.$}
  \label{fig:3Dsims}
\end{figure*}The method as discussed is extensible to 3D in a straightforward manner. Preliminary results so far are essentially identical to the 2D case, with gradual nematic ordering in proportion to the salt concentration, a charge distribution well-approximated as a polynomial function of the order parameter and average structure sizes as predicted by the Ohta-Kawasaki and Onuki models in Eq.~(\ref{eq:biglambda}). A somewhat wider variety of different periodic structures is observed depending on the volume fraction $m$, as shown in Fig.~\ref{fig:3Dsims}. A minority phase tends to form spherical bubbles, which may split or elongate depending on the strength of electrostatic forces. Slightly asymmetric volume fractions lead to tube structures, which may be separate or, for almost symmetric volume fractions, conjoined into a bicontinuous network. Lastly, symmetric volume fractions lead to lamellar structures. The same morphologies as a function of volume fraction have been previously produced by the Ohta-Kawasaki model~\cite{thomas_polymer_2007}. In 2D we observe only lamellar and droplet phases.\par
An interesting avenue of further research lies in the transition region between the droplet and tubular phases. Here we find simulated systems which do not seem to ever converge to a static state, instead exhibiting repeating cycles of droplet nucleation, elongation of the droplet to a tubular shape by electrostatics and eventually splitting and evaporation of the tube as electrostatic pressure builds up. It is possible that a static state would be reached in a less coarse-grained model applying the electrostatic force separately to the two fluid components as we suggested in discussing Fig.~\ref{fig:simulationpsd}b in section \ref{sec:lengths}, but this remains to be seen.
\section{Conclusions}
\label{sec:outlook}
\subsection{Summary}
Based on our simulations we developed a theoretical model giving the charge distribution as a function of the fluid composition at each point in time. When antagonicity is of the order of a few $k_BT$ this function is linear and for higher antagonicities it is a higher-order polynomial. With this model, we can neglect the complicated dynamics of the ions completely and effectively reduce the system from a quaternary mixture to a unary phase field model. We find that electrostatic interactions in the ion dynamics can in many cases be neglected in the parameter space where mesoscopic structure formation happens, as structure formation occurs only when electrostatic ion fluxes are small in comparison with solvation fluxes. The condition of solvation-dominated ion dynamics $\gamma_p>1$, or $\lambda_I\gamma<\epsilon\Delta\mu^2/6e^2$, is identical to the condition of structure formation derived by Onuki~\cite{onuki_solvation_2004}. Assuming for example $\Delta\mu=15k_BT$ at a temperature of $T=330$K and $\epsilon_r=40$ the product of interface width and surface tension has to fulfill $\lambda_I\gamma< 10$pN in order for strong fluid-ion coupling and therewith structure formation to be possible. For most water-oil mixtures this value is much larger under normal conditions but can be expected to rapidly decrease to zero as the temperature approaches the critical point of demixing~\cite{buhn_molecular_2004,pousaneh_molecular_2016}. In a recent paper, Okamoto and Onuki have predicted a similar coupling as we observe of charge and order parameter between nonionic solutes in water-oil mixtures to explain the so-called Ouzo-effect~\cite{okamoto_theory_2018}.\par
By showing equivalency in the case of low antagonicity to the Ohta-Kawasaki model we motivate the observed structure formation and nematic ordering and predict the resulting structure sizes. While it remains to be seen how the scaling laws and morphologies differ from the Ohta-Kawasaki model due to nonlinear coupling of charge and fluid composition for high $\Delta \mu$, we observe excellent agreement with the scaling laws of the Ohta-Kawasaki model up to $\Delta\mu=8k_BT.$ Our 3D simulations show essentially the same morphologies as a function of the volume fraction $m$ as is known from the Ohta-Kawasaki model~\cite{thomas_polymer_2007}. On a similar note, Pousaneh and Ciach recently showed in their study of confined mixtures~\cite{pousaneh_effect_2014} how binary fluid mixtures containing antagonistic salts can also be modeled via the Landau-Brazovskii free energy model, which is recovered by the Ohta-Kawasaki model in the weak-segregation limit~\cite{huang_coarse-grained_2007}.\par
Our results on nematic ordering point towards the possibility of controlling the average size of ordered domains at a given time or at least the speed of nematic ordering via a number of system parameters affecting $\Lambda$, namely the salt concentration $c_s$ but possibly also the temperature, which will strongly affect the ratio of $\lambda_I^3/\gamma$ close to the critical point. It remains to be studied, whether thermal fluctuations eventually stop nematic ordering or it progresses continuously as suggested by our simulations.\par
Comparing to experimental data from SANS we numerically reproduce and explain the dissolution of structures at high salt concentration as a result of remixing caused by high electrostatic fluid forces and zero or negative effective surface tension when $\Lambda>\Lambda_c.$ For equal volume fractions $m=0.5$ and again assuming $\Delta\mu=15k_BT$ and $\epsilon_r=40$ this condition is equivalent to $\lambda_I^3c_s^2/\gamma>1.9\cdot 10^{26}$N$^{-1}$ m$^{-2}$ for a critical $\Lambda_c\approx 3.1.$ We also interpret the secondary peaks observed in the experimental data as a result of reduced fluid and ion concentrations at the interfaces.\par
Future work may include quantifying the effects of dielectric permittivity contrasts between the two fluid components, further testing the applicability of the Ohta-Kawasaki model for high, but still realistic values of $\Delta \mu \approx 15k_BT,$ a more detailed study of the dynamics of nematic ordering, and reproducing the lamellar phase at low volume fractions, as observed in the experiments of Sadakane et al.~\cite{sadakane_membrane_2013}.
\subsection{Acknowledgements}
The authors acknowledge financial support by the Deutsche Forschungsgemeinschaft (DFG) within the Cluster of Excellence "Engineering of Advanced Materials”, the Initiative and Networking Fund of the Helmholtz Association and by the Bavarian Ministry of Economic Affairs and Media, Energy and Technology for the joint projects in the framework of the Helmholtz Institute Erlangen-N{\"u}rnberg for Renewable Energy (IEK-11) of Forschungszentrum J{\"u}lich. We thank the J{\"u}lich Supercomputing Centre and the High Performance Computing Centre Stuttgart for the technical support and the allocated CPU time.
% \bibliography{nematic}

\end{document}